\documentclass[journal,twocolumn]{IEEEtran}
\usepackage{float}
\IEEEoverridecommandlockouts
\usepackage{cite}
\usepackage{stfloats} 
\usepackage{amssymb,amsfonts}
\usepackage{graphicx}
\usepackage{adjustbox}
\usepackage{textcomp}
\usepackage[utf8]{inputenc}
\usepackage[fleqn]{amsmath}
\usepackage{xcolor}
\usepackage{soul}
\usepackage{booktabs}
\usepackage{array}
\usepackage{algorithm}
\usepackage{algcompatible}
\usepackage{algpseudocode}
\usepackage{graphicx}
\usepackage{textcomp}
\usepackage{xcolor}
\usepackage{graphics}
\usepackage{graphicx}
\usepackage{epsfig}
\usepackage{fancyhdr}
\usepackage{enumitem}
\usepackage{wasysym}
\usepackage{latexsym}
\usepackage{color}
\usepackage[caption=false,labelfont=sf,textfont=sf]{subfig}
\usepackage{cite}
\usepackage{mathtools}
\usepackage{comment}
\usepackage{booktabs}
\usepackage{pifont}
\usepackage[hidelinks]{hyperref}  % For clickable references
\usepackage{setspace}
\newcommand{\xmark}{\ding{55}}%
\def\BibTeX{{\rm B\kern-.05em{\sc i\kern-.025em b}\kern-.08em
T\kern-.1667em\lower.7ex\hbox{E}\kern-.125emX}}
\begin{document}
%\doublespacing
\title{Amalgamated CHIRP and  OFDM for ISAC }

\author{ Pankaj Kumar,\textit{ Student Member,~IEEE,} Mohammed~El-Hajjar,\textit{ Senior Member, IEEE,} Ibrahim A.~Hemadeh, \textit{Senior Member, IEEE,} Yasser Mestrah, Suraj Srivastava, \textit{Member, IEEE,}  Aditya~K.~Jagannatham, \textit{ Senior Member, IEEE,} and Lajos~Hanzo, \textit{ Life Fellow, IEEE}

%\thanks{Copyright (c) 2026 IEEE. Personal use of this material is permitted. However, permission to use this material for any other purposes must be obtained from the IEEE by sending a request to pubs-permissions@ieee.org.}
\thanks{The financial support of the following Engineering and Physical Sciences Research Council (EPSRC) projects is gratefully acknowledged: Platform for Driving Ultimate Connectivity (TITAN) under Grant EP/Y037243/1 and EP/X04047X/1; Robust and Reliable Quantum Computing (RoaRQ, EP/W032635/1); India-UK Intelligent Spectrum Innovation ICON UKRI-1859; PerCom (EP/X012301/1); EP/X01228X/1; EP/Y037243/1;}
\thanks{P. Kumar, M. El-Hajjar, and L. Hanzo are with the School of Electronics and Computer Science, University of Southampton, Southampton SO17 1BJ, U.K. (e-mail:Pankaj.Kumar@soton.ac.uk; meh@ecs.soton.ac.uk; lh@ecs.soton.ac.uk). I. Hemadeh and Y. Mestrah are with InterDigital, EC3A 3DH London, U.K. (e-mail: Ibrahim.Hemadeh@InterDigital.com;
 Yasser.Mestrah@InterDigital.com). S. Srivastava is with the Department of Electrical Engineering, Indian Institute of Technology Jodhpur,
342030, India (e-mail: surajsri@iitj.ac.in).
 A. K. Jagannatham is with the Department of Electrical Engineering, Indian Institute of Technology, Kanpur, 208016, India (e-mail:adityaj@iitk.ac.in).}

}
\maketitle
\begin{abstract}
Integrated Sensing and Communication (ISAC) requires the development of a waveform capable of efficiently supporting both communication and sensing functionalities. This paper proposes a novel waveform that combines the benefits of both the orthogonal frequency division multiplexing (OFDM) and the Chirp waveforms to improve both the communication and sensing performance within an ISAC framework. Hence, a new architecture is proposed that utilizes the conventional communication framework while leveraging the parameters sensed at the receiver (Rx) for enhancing the communication performance. We demonstrate that the affine addition of OFDM and chirp signals results in a near constant-envelope OFDM waveform, which effectively reduces the peak-to-average power ratio (PAPR), a key limitation of traditional OFDM systems. Using the OFDM framework for sensing in the conventional fashion requires the allocation of some resources for sensing, which in turn reduces communication performance. As a remedy, the proposed affine amalgam facilitates sensing through the chirp waveform without consuming communication resources, thereby preserving communication efficiency.  Furthermore, a novel technique of integrating the chirp signal into the OFDM framework at the slot-level is proposed to enhance the accuracy of range estimation. The results show that the OFDM signal incorporated with chirp has better autocorrelation properties, improved root mean square error (RMSE) of range and velocity, and lower PAPR. Finally, we characterize the trade-off between communications and sensing performance.
\end{abstract}
\vspace{-1.5em}
\section{Introduction}
Traditionally, communications and radar systems have been designed and operated independently, often using separate spectral resources\cite{10418473}. Modern communication systems face increasing thirst for spectrum and hardware efficiency, particularly in the context of next-generation technologies\cite{11017652}. These challenges underscore the benefits of Integrated Sensing and Communication (ISAC) systems, which integrate sensing and communication functionalities into a shared infrastructure\cite{8869705}. By leveraging the same waveforms and spectrum, ISAC addresses critical issues such as spectrum scarcity \cite{10215321 }, energy consumption \cite{10328645}, while mitigating the cost and complexity of maintaining separate systems~\cite{10565781}.
%\begin{figure}[htbp]
%\centerline{\includegraphics[width=0.70\linewidth]{isac_application.png}}
%  \caption{Illustration for the applications of ISAC.\cite{9705498}}
%\label{fig1}
%\end{figure}

ISAC systems find applications in diverse fields. For example, in autonomous vehicles, ISAC supports radar-based sensing for obstacle detection along with inter-vehicle communication \cite{9947033}. The {International Telecommunication Union Radiocommunication Sector (ITU-R)} has established high-level guidelines for {ISAC} as part of its {IMT-2030 (6G)} framework \cite{11017652}, emphasizing the convergence of communications and sensing functionalities to address emerging needs such as autonomous mobility, environmental monitoring, and smart infrastructure \cite{10529727}. This integration is driven by the potential benefits of ISAC, including efficient spectrum utilization, reduced hardware redundancy, and enhanced capabilities for precise localization and environmental awareness\cite{10608156}. Building on ITU-R’s vision, the {3rd Generation Partnership Project (3GPP)} has undertaken a {Feasibility Study on ISAC} under {Release 19}, detailed in {Technical Report 22.837} \cite{3gpp_tr22837}, to explore practical implementation strategies. These include spectrum allocation, waveform design, and resource management techniques, ensuring ISAC's seamless integration into future cellular networks \cite{10791445,10788035}.

The waveform design is a critical aspect of ISAC systems, since it simultaneously serves as the foundation for achieving efficient communication and accurate sensing \cite{9724187}. ISAC systems must enable high-speed and reliable data transmission, while performing tasks such as range, velocity, and angle estimation. The choice of waveform directly determines the system's ability to meet these dual objectives \cite{10255745}. A well-designed waveform strikes a trade-off between sensing accuracy and communication performance.
In the waveform design of ISAC systems, key properties such as autocorrelation sidelobes \cite{8782136}, correlation peaks \cite{9130747}, and the ambiguity function \cite{doi:https://doi.org/10.1002/0471663085.ch3} play a pivotal role in determining sensing performance. A well-designed waveform should exhibit a sharp main peak in its autocorrelation function with minimal sidelobes to ensure accurate range and velocity estimation while avoiding ambiguity \cite{10771629}. Having a high peak-to-sidelobe ratio (PSLR) is essential for distinguishing closely spaced targets, while properties like range resolution and Doppler tolerance depend on the waveform’s bandwidth and robustness to frequency shifts \cite{5776640}. 

ISAC waveforms, which underpin ISAC technology, have been the subject of comprehensive research. Wei \textit{et al.} \cite{10012421} presented a comprehensive survey and explored a suite of innovations in signal design, processing, and optimization, while addressing challenges and potential applications. The ISAC waveforms are classified primarily into three categories depending on the design methods, such as  communication-centric waveforms (CCW) \cite{5776640,9724198,9743516}, sensor-centric waveforms (SCW) \cite{9285278,8563361,9525432}, and joint waveform optimization and design (JWOD)\cite{7953658,10411942,9652071}.
Communication-centric waveforms, such as OFDM \cite{5776640}, single-carrier waveforms, and spread spectrum techniques \cite{9724198}, relying on PSK/QAM modulated signals, are designed primarily for efficient data transmission, offering high spectral efficiency, robustness to channel impairments, and low bit error rates (BER) \cite{9743516}. Although these waveforms excel in communication tasks, they face significant limitations when harnessed for sensing. For instance, they often exhibit poor autocorrelation properties, leading to high sidelobes that reduce the accuracy of range and velocity estimation, while introducing ambiguities in target detection \cite{https://doi.org/10.1049/iet-rsn.2017.0369}.

The CCW rely on traditional communication waveforms to extract the sensing information of interest from the echoes of signals reflected by nearby objects. In 2011, Sturm \textit{et al.} \cite{5776640} developed a framework that employs OFDM and incorporates advanced sensing signal processing algorithms, with an emphasis on optimizing the overall performance. The accuracy of the estimation depends on the number of symbols allocated for sensing, thereby reducing the number of symbols available for communication. The orthogonal
time frequency space (OTFS) waveform \cite{9904976,10630836, 9903393}, formulated within the delay-Doppler framework is particularly effective in tackling the Doppler effects. The characteristics of OTFS make it eminently suitable for systems that require the integration of communication and sensing capabilities \cite{9903393}.

Furthermore, Sanson \textit{et al.} \cite{8832140} developed a signal based on generalized frequency division multiplexing (GFDM) for ISAC, emphasizing its advantages in terms of reduced inter-user interference and guard band requirements in comparison to OFDM. This approach exhibits notable improvements in range resolution and demonstrates resilience in interference-limited environments. On the other hand, Koslowski \textit{et al.} \cite{6851369}
 explored the adaptation of joint OFDM-based radar and communication systems to Filter Bank Multicarrier (FBMC) signals. They demonstrated that radar imaging methodologies, specifically the two-dimensional periodogram employed for range-Doppler estimation in OFDM radar, can be adapted for FBMC radar systems.
 Additionally, Sangeeta \textit{et al.} \cite{9746551} investigated the use of orthogonal chirp division multiplexing (OCDM) as a waveform, showing that it provides robust communication performance in time-frequency selective channels and achieves both accurate bistatic range as well as Doppler velocity estimation, validating its suitability for automotive ISAC applications. Most communication waveforms can be utilized for sensing applications, provided that the transmit waveform is known to the receiver a priori \cite{8918315}. Hence, the major condition for using the CCW is to \textit{use pilot signals} and it is suitable for monostatic scenarios.
 
By contrast, sensing-centric waveforms are designed for accurate parameter estimation, such as range, velocity, and angle, making them ideal for radar and other sensing applications. Linear frequency modulated (LFM) and chirp waveforms are the most common sensing-centric ones used for conveying communication symbols in ISAC 
systems \cite{9285278,https://doi.org/10.1049/iet-rsn.2017.0369,8563361,9525432}. Both LFM and chirp waveforms offer robust sensing capabilities, but their data rates for communication are constrained. Therefore, the aforementioned methods typically demonstrate improved radar sensing capabilities, albeit at a reduced communication rate. Although information-bearing symbols can be readily inserted into any radar waveform through frequency shifts \cite{https://doi.org/10.1049/iet-rsn.2017.0369,8782136} or phase modulation \cite{https://doi.org/10.1049/iet-rsn.2017.0369,8563361}, this inevitably incurs some degradation in sensing performance, while the hardware architecture of the transceiver will become more complex to the transmission and reception of information data \cite{9130747}.

To improve the sensing performance of OFDM systems, many chirp-based OFDM systems were proposed \cite{https://doi.org/10.1049/iet-rsn.2017.0369,8782136,9130747}. Li \textit{et al.} \cite{https://doi.org/10.1049/iet-rsn.2017.0369} proposed the embedding of phase-modulated communication information into the OFDM chirp waveform for delay-Doppler radar. The method proposed in \cite{https://doi.org/10.1049/iet-rsn.2017.0369} incorporates the chirp at the transmitter (Tx) in the analog domain, which can lead to synchronization problems and also reduce the flexibility in incorporating chirp into OFDM signals. In \cite{8782136}, an innovative approach was proposed where the LFM signal is transformed into the frequency domain using FFT. This transformation effectively mitigates the delay-Doppler coupling, simplifying the signal processing requirements. Furthermore, Zhao \textit{et al.} \cite{9130747} proposed a novel OFDM chirp waveform based design for improving the correlation properties for both radar and communication applications. However, the method faces high computational complexity and requires careful tuning of Pareto weights to balance the conflicting design objectives, which can be challenging in some applications.

Joint waveform design is mainly based on optimizing the desired waveform based on different applications. It hinges on balancing the sensing versus communication performance. There are numerous performance metrics, such as the signal-to-interference plus noise ratio (SINR) \cite{7953658}, mutual interference (MI) \cite{10411942}, Cramer-Rao bound (CRB) \cite{9652071}, of which the waveform can be optimized. In \cite{7953658}, the authors conceived a spectrum-sharing strategy for colocated MIMO radar and communication that maximizes the radar’s SINR, while meeting specific communication rate and power requirements. Additionally, the authors of \cite{10411942} optimize the MI between the response matrix of a target and the reflected signals, while satisfying the data rate requirements of the communications users. Constrained by the SINR requirement of each user and the total transmit power allocation, Liu \textit{et al.} \cite{9652071} minimized the CRB of the beamforming matrix. 
\begin{comment}

 \begin{table*}[h!]
    \centering
    \caption{Contrasting contributions to the existing literature}
\label{tab_lit_rev}

\begin{tabular}{|l|c|c|c|c|c|c|c|c|c|c|c|c|c|c|}
    \hline
 & \cite{5776640}  &  \cite{9904976} &
\cite{9746551} & \cite{8918315} & \cite{9285278} & \cite{{https://doi.org/10.1049/iet-rsn.2017.0369}}
&\cite{9525432} & \cite{8782136} &
\cite{8114253} & \cite{9921271} &

Proposed Work \\ [0.5ex]

\hline
Conception of AAC-OFDM &  & &  &  &
 &  & & &  &  &
\checkmark \\
\hline
Symbol versus Slot-based Chirp   & &  &  &  &  &
&  &  &  &  &
\checkmark  \\
\hline
Ambiguity function of AAC-OFDM & &  &  &  &  & &
 &  &  &  & \checkmark \\
\hline
Multi-path channel  & \checkmark &  &\checkmark  & \checkmark &  &   &  &  & \checkmark & & \checkmark \\
\hline
Chirp based OFDM  & & \checkmark & \checkmark &  & \checkmark 
& \checkmark  & \checkmark  &\checkmark  &  &  & \checkmark \\
\hline
Bistatic ISAC &  &  & \checkmark & \checkmark &\checkmark  &
&  & \checkmark & & &
\checkmark \\
\hline
Monostatic ISAC & \checkmark &\checkmark  &  & & & \checkmark & \checkmark & &\checkmark
 & \checkmark & \checkmark\\
\hline
%Robust design &  &  &  &  &  &   &  &  &  & & \checkmark\\
%\hline
\end{tabular}
\end{table*}
\end{comment}
\begin{table*}[t]
    \centering
    \caption{Contrasting contributions to the existing literature}
    \label{tab_lit_rev}

    % ↓↓↓ Height/spacing tweaks (local to this table) ↓↓↓
    \small                                 % or \footnotesize / \scriptsize
    \setlength{\tabcolsep}{4pt}            % default ~6pt; smaller packs columns (may help wrapping)
    { \renewcommand{\arraystretch}{0.9}    % < 1.0 shrinks row height; try 0.9–0.95
    \begin{tabular}{|l|c|c|c|c|c|c|c|c|c|c|c|}
        \hline
        & \cite{5776640} & \cite{9904976} & \cite{9746551} & \cite{8918315} & \cite{9285278} 
        & \cite{https://doi.org/10.1049/iet-rsn.2017.0369}
        & \cite{9525432} & \cite{8782136} & \cite{8114253} & \cite{9921271} & Proposed Work \\
        \hline
        Conception of AAC-OFDM           &  &  &  &  &  &  &  &  &  &  & \checkmark \\
        \hline
        Symbol versus Slot-based Chirp    &  &  &  &  &  &  &  &  &  &  & \checkmark \\
        \hline
        Ambiguity function of AAC-OFDM    &  &  &  &  &  &  &  &  &  &  & \checkmark \\
        \hline
        Multi-path channel                & \checkmark &  & \checkmark & \checkmark &  &  &  &  & \checkmark &  & \checkmark \\
        \hline
        Chirp based OFDM                  &  & \checkmark & \checkmark &  & \checkmark & \checkmark & \checkmark & \checkmark &  &  & \checkmark \\
        \hline
        Bistatic ISAC                     &  &  & \checkmark & \checkmark & \checkmark &  &  & \checkmark &  &  & \checkmark \\
        \hline
        Monostatic ISAC                   & \checkmark & \checkmark &  &  &  & \checkmark & \checkmark &  & \checkmark & \checkmark & \checkmark \\
        \hline
    \end{tabular}
    } % end local arraystretch
\end{table*}

The deployment of ISAC in future wireless networks can be achieved cost-effectively by modifying the structure of existing standardized communication frames for sensing applications \cite{8917703,9921271}, while using bespoke physical layer reference signals for sensing purpose, such as demodulation reference signals (DMRS), phase tracking reference signals (PTRS), channel state information reference signals (CSI-RS), and positioning reference signals (PRS). Wei \textit{ et al.} \cite{9921271}  utilized PRS for estimating both the range and velocity. However, the performance of PRS is highly dependent on the fraction of resources allocated. For instance, allocating resources having a comb size of 2 would use 50\% of the total subcarriers, which would halve the communications capacity.

Against this backdrop, we propose a new waveform that is capable of simultaneously transmitting communication data along with performing sensing tasks without the need for pilot subcarriers. The contributions of this paper are boldly and explicitly contrasted to the literature in Table \ref{tab_lit_rev} and can be summarized as follows:
 \begin{itemize}
    \item We propose an innovative waveform design that employs the affine addition of chirp and OFDM (AAC-OFDM) signals in the time domain (TD). We design a systematic technique of optimizing the affine addition parameters, carefully balancing the trade-off between robust communication performance and precise sensing capabilities. 
    \item We derive the ambiguity function for the AAC-OFDM waveform, which establishes a theoretical framework for evaluating its sensing and communication performance. This derivation serves not only as a foundational tool for optimizing the AAC-OFDM waveform but also underscores its applicability across a range of ISAC scenarios. By formalizing this relationship, the study paves the way for the systematic performance enhancement of the proposed system.
   \item The simulation-based sensing performance results have shown that AAC-OFDM does not require pilot subcarriers or prior knowledge of the transmitted data for performing sensing tasks, which makes it an eminently suitable candidate for enhancing the spectral efficiency and sensing performance of ISAC systems. This property of AAC-OFDM is beneficial in the case of bistatic/multistatic sensing, where the receiver is located at a different base station and does not have prior knowledge of the transmitted bits. This approach provides a practical framework for advancing waveform design, demonstrating its potential to meet the growing demands of ISAC applications.
\begin{comment}
\item The implementation of the {CM-OFDM} waveform has predominantly been explored in the analog domain in existing literature, leading to challenges such as synchronization mismatches and increased error rates. Our work introduces a novel approach by implementing the CM-OFDM and AAC-OFDM waveform entirely in the discrete domain, leveraging advanced digital signal processing techniques. This discrete-domain implementation mitigates mismatches that occur during the signal processing stage, ensuring precise alignment and reducing errors caused by imperfections in the combination process. Furthermore, it provides enhanced control over system operations, allowing for precise waveform design and optimization. This advancement significantly improves the robustness and reliability of ISAC systems, marking a crucial step forward in the practical deployment of CM-OFDM and AAC-OFDM waveforms.
\end{comment}
\item We present a comparative analysis to evaluate the performance of the chirp multiplied OFDM (CM-OFDM) waveform conceived against OFDM used as a position reference signal (PRS) commonly employed in OFDM-aided ISAC systems. Through the ambiguity function, we have shown that the CM-OFDM waveform significantly outperforms OFDM in terms of its range resolution, while maintaining comparable communications performance. This shows that the performance of PRS is improved by CM-OFDM. 
\item We present a novel comparison of the performance of AAC-OFDM and CM-OFDM signals implemented at both the 5G New Radio (NR) slot and symbol levels. The analysis focuses on the root mean square error (RMSE) of range and velocity estimation under each implementation strategy. The results reveal that the slot-level superposition of chirp signals consistently outperforms its symbol-level implementation, offering superior sensing accuracy and robustness. This finding highlights the advantage of slot-level integration in enhancing system performance, providing valuable insights for optimizing the waveform design of ISAC systems. 
\begin{comment}
The study marks a significant contribution by demonstrating how implementation strategies can influence the effectiveness of chirp-superimposed OFDM waveforms and how it can outperform the current OFDM waveform for sensing purposes.
\end{comment}
\end{itemize}
\begin{comment}
\begin{figure}[htbp]
\centering
\includegraphics[height=0.2\textheight, width=1.00\linewidth]{Pictures/mywork/time_domain.eps} % Adjust height and width independently
\caption{Time domain structure of 5G NR.}
\label{fig:time_domain}
\end{figure}

\begin{figure}[htbp]
\centering
\includegraphics[height=0.2\textheight, width=0.7\linewidth]{Pictures/mywork/time_frequncy.eps} % Adjust height and width independently
\caption{Time frequency structure of OFDM.}
\label{fig:frequency_domain}
\end{figure}
\end{comment}
%\subsection{Notations}

The rest of the paper is organized as follows. In Section \ref{sec:system_model}, we discuss the system model along with the channel model used in this paper. In Section~\ref{sec:cm_aac_ofdm}, we formulate the waveforms of AAC-OFDM and CM-OFDM, followed by analysis of both waveforms in Section~\ref{sec:analysis}. In Section~\ref{sec:slot_vs_symbol}, we proposed the implementation technique to improve the sensing performance, followed by their simulation-based characterization for communication and sensing tasks in Section~\ref{sec:simulation_results}. Our conclusions are provided in Section~\ref{sec:conclusion}.

Notation: Vectors and matrices are denoted by boldfaced lower and upper case letters, respectively, while scalars are denoted by lower case letters. Furthermore, $\mathbf{x} \sim$ $\mathcal{C}\mathcal{N}(\mathbf{b},\mathbf{A})$ denotes a complex Gaussian random vector with mean $\mathbf{b}$ and covariance matrix $\mathbf{A}$. $\mathbf{I}$ represents an identity matrix of suitable dimension. 
\begin{figure}[htbp]
\centerline{\includegraphics[width=0.70\linewidth,height=2.8cm]{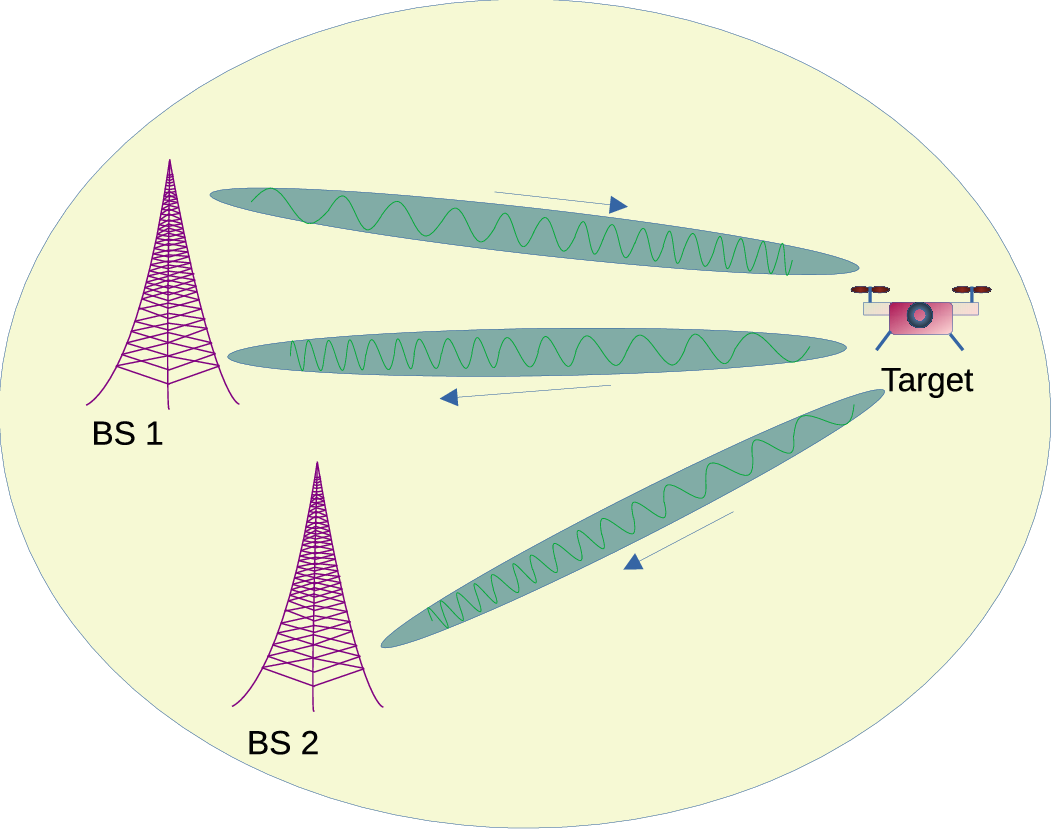}}
\caption{System model for ISAC based on AAC-OFDM and CM-OFDM.}
\label{fig:system_model}
\end{figure}
\vspace{-1.6em}
\section{System Model}\label{sec:system_model}
The system model shown in Fig.~\ref{fig:system_model} represents both monostatic and bistatic sensing. Monostatic sensing is established from a distributed monostatic system by arranging the virtual co-location of the transmitter and receiver upon connecting them via fiber. Both the transmitter and receiver perfectly know the transmitted signals using this setup. By contrast, bistatic sensing has the transmitter and receiver at separate locations. As a result, the receiver may only have partial knowledge of the transmitted signals \cite{10437391}.

Consider a single-input and single-output (SISO) system, where the distance between the base station (BS) and the objects is $R$. The BS 1 in Fig.~\ref{fig:system_model} transmits a signal that is reflected by the objects, and then the receiver at BS 1 or BS 2, depending on the type of sensing processes, extracts the necessary information about the object's location.
\begin{figure}[htbp]
\centerline{\includegraphics[width=0.9\linewidth,height=2.9cm]{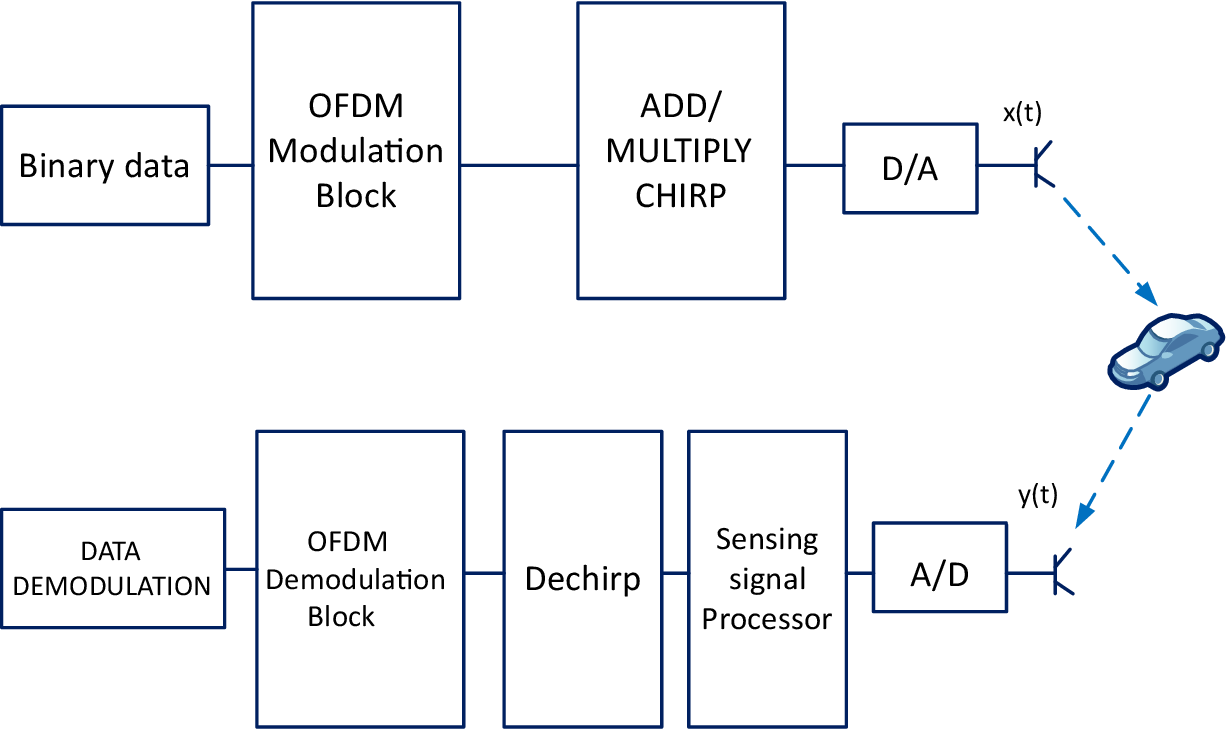}}
\caption{Block diagram of the CM-OFDM and AAC-OFDM system architectures.}
\label{fig:block_diagram}
\end{figure}
\vspace{-0.1\baselineskip}
The block diagram provided in Fig. \ref{fig:block_diagram} illustrates the system architecture intended for ISAC applications. In the transmission chain, binary data are initially processed through an OFDM block, then the chirp signal is incorporated in the resultant OFDM signal for supporting the ISAC functionality. The processed signal is then transformed into an analog format using a digital-to-analog (D/A) converter and delivered across the channel as \textit{x}(t). Upon reception, the received signal \textit{y}(t) is subjected to analog-to-digital (A/D) conversion to enable digital processing. The sensing signal processor in Fig.~\ref{fig:block_diagram} analyzes the received signal for extracting the sensing-related information, including range and velocity estimation.
  
In much of the literature, the sensing and communication signals are typically processed jointly at the receiver. By contrast, the proposed system extracts the sensing information separately and uses it for enhancing communication performance. By embedding the sensing signal within the communication framework, the receiver leverages the sensing information extracted for optimizing communication. The dechirping operation in Fig.~\ref{fig:block_diagram} is utilized for eliminating the chirp component. The signal subsequently traverses the OFDM block, where it undergoes demodulation using FFT for retrieving the transmitted data. Ultimately, the data demodulation block in Fig.~\ref{fig:block_diagram} extracts the original binary information.
  
Two distinct systems have been conceived in this work. One is formulated by the affine addition of chirp to OFDM termed as AAC-OFDM, while the other is formulated by the multiplicative amalgam of chirp with OFDM, referred to as CM-OFDM. The delay-Doppler channel impulse response is modeled as a sum of multiple propagation paths \cite{8424569}:
\begin{IEEEeqnarray}{c}
h(\tau, f) = \sum_{i=0}^{P-1} h_i\, \delta(\tau-\tau_i)\, \delta(f-f_{d,i}),
\end{IEEEeqnarray}
where \(P\) is the total number of propagation paths and each path \(i\) has its own complex gain \(h_i\), delay \(\tau_i\), and Doppler shift \(f_{d,i}\). Finally, $\delta(\cdot)$ denotes the Dirac delta function.
\begin{comment}
Accordingly, the channel is simplified to a single-path model:
\begin{IEEEeqnarray}{c}
h(\tau, \nu) = h_0\, \delta(\tau-\tau_0)\, \delta(\nu-\nu_0),
\end{IEEEeqnarray}
where \(h_0\) is a constant gain (with no additional phase contribution from the channel aside from the Doppler effect), \(\tau_0\) is the LOS delay, and \(\nu_0\) is the Doppler shift associated with the LOS path.
\end{comment}
 
\section{Proposed waveform based on OFDM with Pulse compression radar systems}\label{sec:cm_aac_ofdm}
In this section, the waveforms of AAC-OFDM and CM-OFDM will be discussed, highlighting their characteristics and applications. The AAC-OFDM waveform is our novel proposed approach where the OFDM and chirp signals are combined using an affine addition method. This design aims to decouple the sensing and communication functionalities, allowing a flexible ISAC waveform design. By contrast, in CM-OFDM  the chirp's linear frequency modulation is exploited for sensing, while the underlying OFDM structure facilitates communication. \vspace{-1em}  
\subsection{{AAC-OFDM}}
The discrete-time domain representation of the AAC-OFDM signal $ {a}(l)$ is defined as 
\begin{IEEEeqnarray}{c}
{a}(l) = (1-\alpha) {s}(l) + (\alpha) {c}(l),
\label{eq:aac-ofdm}
\end{IEEEeqnarray}
where $\alpha$ is the power allocation factor, while ${s}(l)$ and ${c}(l)$ are the OFDM signal and chirp signal, respectively, defined as
\begin{IEEEeqnarray}{c}
{s}(l) =  \sum_{m=0}^{M-1}s_{m}(l)\, \text{rect}\left( \frac{l - mN}{N} \right),
\label{eq:ofdm_signal_form}
\end{IEEEeqnarray}
\begin{IEEEeqnarray}{c}
{c}(l) =  \sum_{m=0}^{M-1}c_{m}(l)\, \text{rect}\left( \frac{l - mN}{N} \right),
\label{eq:chirp_signal_form}
\end{IEEEeqnarray}
where we have:
\[
\text{rect} \left( \frac{l - mN}{N} \right) =
\begin{cases}
1, & mN \leq l \leq (m+1)N-1, \\
0, & \text{otherwise}.
\end{cases}
\]
The time index \( l \) represents discrete sampling points within each symbol, ranging from \( l =mN \) to \( l = (m+1)N-1 \), facilitating the time-domain description of the signal. Furthermore, \( s_m(l) \) represents samples of the $m_{th}$ OFDM symbol after the $N$-point IFFT operation applied to $N$ number of sub-carriers, which can be represented as:
 \begin{IEEEeqnarray}{c}
s_{m}(l) = \sum_{n=0}^{N-1} X_{m}(n) e^{j \frac{2\pi nl}{N}}.
\label{eq:ofdm_signal_form_iift}
\end{IEEEeqnarray}
The modulated data symbol \( X_{m}(n) \) conveys the actual information payload for the \( n_{th} \) subcarrier during the \( m_{th} \) symbol duration. Additionally, \( c_m(l) \) represents the chirp which is applied to the $m_{th}$ symbol. The chirp signal parameters are perfectly aligned with the OFDM parameters, ensuring the incorporation of chirp results in a seamlessly integrated system. This design ensures that changes in the key OFDM parameters, such as the subcarrier spacing or the number of subcarriers, automatically adjust the chirp's time duration and all other associated parameters, maintaining coherence and compatibility between the two waveforms. The chirp signal  \( c_m(l) \) is defined in the discrete time domain as  
\begin{IEEEeqnarray}{rCl}
c_{m}(l) &=& \sum_{n=0}^{N-1} q_{m}(n)\,
e^{j \pi \beta_{m,n} \bigl(T_{s}l \bigr)^{2}},
\label{eq:chirp_signal_l_and_m}
\end{IEEEeqnarray}
where \( q_{m}(n) \) represents the amplitude of the chirp corresponding to the \( n_{th} \) subcarrier and \( m_{th} \) symbol, capturing the signal's strength for each subcarrier and symbol. The parameter \( \beta_{m,n} \) denotes the chirp rate associated with the \( n_{th} \) subcarrier and \( m_{th} \) symbol, determining the frequency modulation of the chirp. The sampling interval \( T_{s} = \frac{1}{N \Delta f} \) is the reciprocal of the product of the total number of subcarriers \( N \) and the subcarrier spacing \( \Delta f \), ensuring appropriate sampling of the signal in the time domain.
The chirp rate for the chirp signal is defined according to the OFDM symbol and can be denoted as follows:
\begin{IEEEeqnarray}{c}
\beta_{\text{m}, n_{j,i}} = \frac{(n_j - n_i) \Delta f}{T_{\text{c}}},
\label{eq:beta_max}
\end{IEEEeqnarray}
where \( n_j \) and \( n_i \) represent the indices of the OFDM subcarriers, while for the chirp signal, it represents the start and end frequency of the chirp. The \( (n_j - n_i)\Delta f \) in \eqref{eq:beta_max} sweeps the total bandwidth \(B_c\) of the chirp. Furthermore,  \(T_c\) is the chirp duration, which is adjusted according to the OFDM symbol duration \(T_o\). If the chirp is imposed on a single OFDM symbol, then  \(T_c=T_o\). We have \eqref{eq:beta_max} highlighting how the chirp's rate depends on the difference between the subcarrier indices, scaled by the duration of the OFDM symbol.

This demonstrates that the chirp spans a frequency range proportional to the subcarrier spacing, normalized by the OFDM symbol duration. The frequency ramp created by the chirp naturally aligns with the frequency domain (FD) structure of the OFDM subcarriers, allowing the chirp to be integrated into the system. This integration enables the chirp to encode the subcarrier frequency differences, while enhancing the signal's utility for both sensing and communication applications. 

In our analysis, we redefine 
\(
\beta_{\text{m}, n_{j,i}}
\)
as 
\(
\beta_{\text{m}, n},
\)
which implies that for each fixed index \( n \), the indices are set for ensuring that 
\(
n_i = n\) and \(n_j = n+1
\).
We substitute the expressions for \( s_{m}(l) \) and \( c_m(l) \) from \eqref{eq:ofdm_signal_form_iift} and \eqref{eq:chirp_signal_l_and_m} into \eqref{eq:ofdm_signal_form} and \eqref{eq:chirp_signal_form}, respectively.
Then we can substitute \eqref{eq:ofdm_signal_form} and \eqref{eq:chirp_signal_form} into \eqref{eq:aac-ofdm}, where the AAC-OFDM signal ${a}(l)$ is represented as
\begin{equation}
\begin{aligned}
   {a}(l) &= \sum_{m=0}^{M-1}  \sum_{n=0}^{N-1} \Big[ 
    (1-\alpha) X_{m}(n) e^{j \frac{2\pi nl}{N}} \, +\\
    &\quad  \alpha q_{m}(n) e^{j \pi \beta_{m,n} \left( T_{\text{s}}l \right)^2} 
    \Big] \, \mathrm{rect} \left( \frac{l - mN}{N} \right).
\end{aligned}
\label{eq:a_m_multiline}
\end{equation}
The transmitted signal $a(t)$ of AAC-OFDM after processing by the DAC block in Fig. \ref{fig:block_diagram} is given by:
\begin{multline}
a(t) = \sum_{m=0}^{M-1} \sum_{n=0}^{N-1} (1-\alpha) X_{m}(n) e^{j2\pi n\Delta f t} 
\text{rect}\left(\frac{t-mT_{\text{o}}}{T_{\text{o}}}\right)\\
+ \sum_{m=0}^{M-1} \sum_{n=0}^{N-1} \alpha q_{m}(n) e^{j\left(\pi \beta_{m,n} (t)^2\right)} 
\text{rect}\left(\frac{t-mT_{\text{o}}}{T_{\text{o}}}\right),
\label{eq:a_t}
\end{multline}
where \( \mathrm{rect}(t/T_{\text{o}}) \) is the rectangular function, defined as:
\[
\mathrm{rect}\left(\frac{t}{T_{\text{o}}}\right) =
\begin{cases} 
1, & 0 \leq t < T_{\text{o}}, \\
0, & \text{otherwise}.
\end{cases}
\]
Furthermore, \( T_{\text{o}} \)=\( T \)+\( T_{\text{CP}} \) is the duration of the total symbol, defined by the subcarrier spacing \( \Delta f \) as \( T = \frac{1}{\Delta f} \), where \( T_{\text{CP}} \) is the cyclic prefix.

The main advantage of the AAC-OFDM signal is that at the receiver, the knowledge of chirp rate is sufficient for estimating the range and velocity of the target object, which is an advantage in the bistatic scenario. The AAC-OFDM received signal can be written as
\begin{equation}
\begin{aligned}
    r(t) &= \sum_{m=0}^{M-1} \sum_{n=0}^{N-1}\sum_{i=0}^{P-1}\xi_{i}(1-\alpha) X_{m}(n) 
    e^{j2\pi n\Delta f \left(t - \frac{2R_i}{c}\right)} \\
    &\quad \times \text{rect}\left(\frac{t - mT_{\text{o}} - \frac{2R_i}{c}}{T_{\text{o}}}\right)e^{j2\pi f_{d,i} t} \\
    &\quad + \sum_{m=0}^{M-1} \sum_{n=0}^{N-1}\sum_{i=0}^{P-1}\xi_{i}\alpha q_{m}(n) 
    e^{j\pi \beta_{m,n} \left(t - \frac{2R_i}{c}\right)^2} \\
    &\quad \times \text{rect}\left(\frac{t - mT_{\text{o}} - \frac{2R_i}{c}}{T_{\text{o}}}\right) 
    e^{j2\pi f_{d,i} t}+n(t).
\end{aligned}
\label{eq:r_t_one_column}
\end{equation}
 The received AAC-OFDM signal is influenced by several critical factors. The amplitude scaling factor \(\xi_{i} \) accounts for the effects of target reflectivity, attenuation, and complex channel gain \(h_i\) of the path, while \(n(t)\) is the additive white Gaussian noise (AWGN). The term \( \frac{2R_i}{c} \) represents the round trip propagation delay, where \( R_i \) is the distance covered by the $i_{th}$ path after being reflected from the object and \( c \) is the speed of light, ensuring that the timing of the signal accurately reflects the range to the object. The Doppler shift \( f_{d,i} \) arises from the relative motion between the transmitter and the object, causing a frequency shift in the signal. The velocity of the object is known from the Doppler frequency as $v_i=f_{d,i}c/2f_{c}$, where \( f_c \) is the carrier frequency of the signal. 
Additionally, the rectangular function \( \mathrm{rect} \left( \frac{t - m T_{\text{o}} - \frac{2R_i}{c}}{T_{\text{o}}} \right) \) confines the signal to the duration of the OFDM symbol, while accounting for the propagation delay. \vspace{-1.6em}
\subsection{ CM-OFDM }
The CM-OFDM signal ${k}(l)$ is a multiplicative combination of the OFDM signal and the chirp signal, and is defined as:
\begin{IEEEeqnarray}{rCl}
    {k}(l) &=& \sum_{m=0}^{M-1}\sum_{n=0}^{N-1} 
    X_{m}(n) q_{m}(n)   
    e^{j \left( \frac{2\pi nl}{N} + \pi \beta_{m,n} \left( T_{\text{s}}l \right)^2 \right)} \, \nonumber\\
&&\quad\times  \mathrm{rect}\left(\frac{l - mN}{N}\right).
\label{eq:k_m}
\end{IEEEeqnarray}
The transmitted CM-OFDM signal $k(t)$ after processing by the DAC block in Fig.~\ref{fig:block_diagram} is given as:
\begin{IEEEeqnarray}{rCl}
k(t) &=& \sum_{m=0}^{M-1} \sum_{n=0}^{N-1} X_{m}(n) q_{m}(n) 
e^{j2\pi n\Delta f t + j\pi \beta_{m,n} t^2}  \, \nonumber\\
&&\quad\times 
\mathrm{rect}\left(\frac{t - mT_{\text{o}}}{T_{\text{o}}}\right).
\label{eq:k_t}
\end{IEEEeqnarray}
The CM-OFDM receiver needs prior information of the transmitted data along with the chirp rate for estimating the range and velocity of the target object. Upon transmitting a signal from BS 1 and receiving it
at the receiver of BS 1 as shown in Fig.~\ref{fig:system_model}, the CM-OFDM signal reflected from the  target is described by $y(t)$ as:
\begin{IEEEeqnarray}{rCl}
y(t) &=& \sum_{m=0}^{M-1} \sum_{n=0}^{N-1}\sum_{i=0}^{P-1}\xi_{i} \,q_{m}(n) X_{m}(n) \,e^{j2\pi n\Delta f \left(t - \frac{2R_i}{c} \right)}\,e^{j2\pi f_{d,i} t} \nonumber\\
&&\times e^{j\pi \beta_{m,n} \left( t - \frac{2R_i}{c} \right)^2}\mathrm{rect}\left(\frac{t - mT_{\text{o}} - \frac{2R_i}{c}}{T_{\text{o}}}\right) 
+n(t).
\label{eq:y_t}
\end{IEEEeqnarray}

The sensing depends on the parameters of the chirp sequence introduced into the OFDM signal, and its specific parameters directly predetermine the key metrics such as the achievable range resolution \(\Delta R\), maximum unambiguous range $R_{\text{max}}$, and the maximum unambiguous velocity \(v_{\text{max}}\) formulated as \cite{10371091}
\begin{IEEEeqnarray}{c}
    \Delta R = \frac{c}{2B_c}, \quad R_{\text{max}} = \frac{c f_s T_c}{4B_c},\quad \lvert v_{\text{max}} \rvert < \frac{\lambda}{4 T_{c}},
      \label{eq:range_reso_max}
\end{IEEEeqnarray}
where \( B_c \) is the bandwidth of the chirp, and \( \lambda = \frac{c}{f_c} \) is the wavelength of the signal. From \eqref{eq:range_reso_max} we can conclude that improving the range resolution requires increasing the chirp bandwidth, but this also limits the maximum unambiguous range. However, extending the duration  \(T_c\) of the chirp, can increase the maximum range. On the other hand, a longer \(T_c\) reduces the maximum unambiguous velocity. To solve this design dilemma, we introduce an innovative implementation technique compatible with the 5G New Radio (NR) frame structure \cite{3gpp_38_211}, which will improve all the metrics and enhance the sensing performance. 
\vspace{-0.4em}
\section{Analysis of AAC-OFDM and CM-OFDM}\label{sec:analysis}
%\vspace{-0.1em}
\subsection{Ambiguity Function }\label{sec:ambiguity_function}
The ambiguity function is an important tool for sensing waveform design and analysis, since it characterizes the features of a waveform and its corresponding matched filter. By analyzing the ambiguity function of the transmitted waveform, one can determine the sensing system's resolution capability, measurement accuracy, and ambiguity when optimal matched filtering is employed. The ambiguity function is defined \cite{doi:https://doi.org/10.1002/0471663085.ch3} as :
\begin{IEEEeqnarray}{c}
\chi(\tau, f_d) = \int_{-\infty}^{+\infty} x(t) x^*(t - \tau) e^{j 2\pi f_d t} \, dt,
\label{eq:ambiguity_formula}
\end{IEEEeqnarray}
where \(x(t)\) represents the  transmitted signal, \(\tau\) is the time delay and \(f_d\) is the Doppler frequency shift.
\vspace{-0.6em}
\subsection{Ambiguity Function of AAC-OFDM}
\vspace{-0.4em}
The ambiguity function \( \chi(\tau, f_d) \) of AAC-OFDM is defined as
\begin{IEEEeqnarray}{c}
\chi(\tau, f_d) = \int_{-\infty}^{+\infty}a(t) a^*(t - \tau) e^{j 2 \pi f_d t} \, dt,
\end{IEEEeqnarray}
where \( a(t) \) is the transmitted signal of AAC-OFDM defined in \eqref{eq:a_t}.
Expanding \( a(t) a^*(t - \tau) \) and assuming that equal power is allocated to both the OFDM and to the chirp waveform will results into four terms:
\begin{IEEEeqnarray}{c}
\chi(\tau, f_d) = I_1 + I_2 + I_3 + I_4,
\label{eq:ambigity_all}
\end{IEEEeqnarray}
where we have:
\begin{equation}\label{eq:I1}
I_1 = \int_{-\infty}^{+\infty} c(t)\,c^{*}(t-\tau)\,e^{j2\pi f_d t}\,dt .
\end{equation}
\begin{equation}\label{eq:I2}
I_2 = \int_{-\infty}^{+\infty} s(t)\,c^{*}(t-\tau)\,e^{j2\pi f_d t}\,dt .
\end{equation}
\begin{equation}\label{eq:I3}
I_3 = \int_{-\infty}^{+\infty} c(t)\,s^{*}(t-\tau)\,e^{j2\pi f_d t}\,dt .
\end{equation}
\begin{equation}\label{eq:I4}
I_4 = \int_{-\infty}^{+\infty} s(t)\,s^{*}(t-\tau)\,e^{j2\pi f_d t}\,dt .
\end{equation}

Here, \( I_1 \) represents the ambiguity function of the chirp signal in which \(c(t)\) is the chirp-modulated signal:
\begin{IEEEeqnarray}{rcl}
c(t) = \sum_{m=0}^{M-1} \sum_{n=0}^{N-1} q_{m,n} e^{j \pi \beta t^2} \text{rect} \left( \frac{t - mT_{o}}{T_{o}} \right).
\label{eq:chirp_time}
\end{IEEEeqnarray}
Here we assume that the same chirp rate is used for all chirp signals; hence \(\beta_{m,n}=\beta\).
Substituting \(c(t)\) and \(c^*(t - \tau)\) into the \eqref{eq:I1}, we get
\begin{equation}
\begin{aligned}
     I_1 &= \sum_{m,n} \sum_{m',n'} q_{m,n} q_{m',n'}^* 
    e^{-j \pi \beta \tau^2} \int_{-\infty}^{+\infty} e^{j ( 2 \pi f_d + 2 \pi \beta \tau)t} \\
    &\quad \times \text{rect} \left( \frac{t - mT_{\text{o}}}{T_{\text{o}}} \right) 
    \text{rect} \left( \frac{t - \tau - m'T_{\text{o}}}{T_{\text{o}}} \right) dt.
    \label{eq:chirp_amb_form}
\end{aligned}
\end{equation}
The product of the rectangular windows restricts \(t\) to the overlap:
\begin{IEEEeqnarray}{rCl}
t_{\text{min}} &=& \max\bigl(m\,T_{o},\,\tau + m'\,T_{o}\bigr), \nonumber\\
t_{\text{max}} &=& \min\bigl(m\,T_{o} + T_{o},\,\tau + m'\,T_{o} + T_{o}\bigr).
\label{eq:tmin_max}
\end{IEEEeqnarray}
The effective integration range is \(t \in [t_{\text{min}}, t_{\text{max}}]\). Let us define
\begin{IEEEeqnarray}{rcl}
T_d = \frac{t_{\text{max}} - t_{\text{min}}}{2}, \quad T_a = \frac{t_{\text{max}} + t_{\text{min}}}{2}.
\label{eq:ta_d}
\end{IEEEeqnarray}
Then we can write \eqref{eq:chirp_amb_form}:
\begin{IEEEeqnarray}{rcl}
\begin{aligned}
    I_1(\tau, f_d) &= \sum_{m,n} \sum_{m',n'} 2 T_d q_{m,n} q_{m',n'}^* e^{-j \pi \beta \tau^2} 
     \\
    &\quad \times \text{sinc} \left( 2 \pi(f_d + \beta \tau) T_d \right)  e^{j (2 \pi f_d + 2 \pi \beta \tau) T_a}.
    \label{eq:ambig_chirp}
\end{aligned}
\end{IEEEeqnarray}

The rectangular windows $[mT_o,\,mT_o+T_o]$ and $[\tau+m'T_o,\,\tau+m'T_o+T_o]$ overlap iff
$|\tau+(m'-m)T_o|\le T_o$. Accordingly, in all sums over $m,m'$ we evaluate

\setcounter{equation}{27}

\begin{IEEEeqnarray}{rCl}
T_d &=& \dfrac{T_o - \left|\tau + (m' - m)T_o\right|}{2},
\IEEEyesnumber\label{eq:Td}\IEEEeqnarraynumspace\\
T_a &=& mT_o + \dfrac{ T_o + \tau + (m' - m)T_o}{2}.
\IEEEyesnumber\label{eq:Ta}\IEEEeqnarraynumspace
\end{IEEEeqnarray}
These expressions hold when the relative shift satisfies
\(
\left| \tau + (m' - m)T_o \right| \leq T_o,
\)
otherwise there is no overlap, and the corresponding integral equals zero.
\noindent When we consider a single OFDM symbol, we set $m'=m$ and adopt a symbol-centered time origin. For $|\tau|\le T_o$ this gives
\begin{IEEEeqnarray}{rcl}
T_d = \frac{T_{o} - |\tau|}{2}, \quad T_a = \frac{T_{o} + \tau}{2}.
\label{eq:case_of_tau2}
\end{IEEEeqnarray}
Furthermore, \( I_2 \) in \eqref{eq:ambigity_all} is the cross-ambiguity between the OFDM and chirp signal, which is represented by
\begin{IEEEeqnarray}{rcl}
\begin{aligned}
    I_2 &= \sum_{m,n} \sum_{m',n'} 
   X_{m}(n) q_{m',n'}^*  
    \int_{-\infty}^{+\infty} e^{j 2 \pi n \Delta f t} 
    e^{- j \pi \beta (t - \tau)^2} e^{j 2 \pi f_d t} \\
    &\quad \times \text{rect} \left( \frac{t - m T_o}{T_o} \right)
    \text{rect} \left( \frac{t - \tau - m' T_o}{T_o} \right) dt.
\end{aligned}
\end{IEEEeqnarray}
\vspace{-0.6em}
Applying the change of variables \( u = t - \tau \), so that \( t = u + \tau \) and \( dt = du \), we can rewrite \(I_2\) as
\begin{align}
    I_2 &= \sum_{m,n} \sum_{m',n'} X_{m}(n) q_{m',n'}^* 
    e^{j (2 \pi n \Delta f + 2 \pi f_d) \tau} \notag \\
    &\quad \times \int_{-\infty}^{+\infty}e^{j (2 \pi n \Delta f + 2 \pi f_d) u} 
    e^{- j \pi \beta u^2} \notag \\
    &\quad \times \text{rect} \left( \frac{u + \tau - m T_o}{T_o} \right)
    \text{rect} \left( \frac{u - m' T_o}{T_o} \right) du.
    \label{eq:i_2}
\end{align}
By defining 
\begin{IEEEeqnarray}{rCl}
u_{\min} &=& \max\bigl(m\,T_o - \tau,\;m'\,T_o\bigr), \\ 
u_{\max} &=& \min\bigl((m+1)\,T_o - \tau,\;(m' + 1)\,T_o\bigr),
\label{eq:u_bounds}
\end{IEEEeqnarray}
and using the variable substitution of \(u = \frac{x}{\beta}\),
the integral in \eqref{eq:i_2} can be expressed in terms of Fresnel integrals as follows:
\begin{equation}
\begin{aligned}
    I_2 &= \sum_{m,n} \sum_{m',n'} X_{m}(n) q_{m',n'}^{*} 
    e^{j\kappa \tau} \frac{1}{j\pi \beta} e^{-j \frac{\kappa^2}{4\pi \beta}} \\
    &\; \times \big[ C(\beta u_{\max}) - C(\beta u_{\min}) 
    - j (S(\beta u_{\max}) - S(\beta u_{\min})) \big],
\end{aligned}
\end{equation}
where $\kappa = 2\pi n \Delta f + 2\pi f_d$ and the Fresnel integrals are defined {\cite{ZaghloulAlrawas2024}} as:
\begin{equation}
\begin{aligned}
C(x) = \int_{0}^{x} \cos \left( \frac{\pi t^2}{2} \right) \, dt
,\, S(x) = \int_{0}^{x} \sin \left( \frac{\pi t^2}{2} \right) \, dt.
\end{aligned}
\end{equation}
Still elaborating on \eqref{eq:ambigity_all}, \( I_3 \) is the cross-ambiguity between the chirp signal and the OFDM signal, which can be written as:
\begin{IEEEeqnarray}{rCl}
I_3 &=& \sum_{m,n} \sum_{m',n'} X_{m}^*(n) \, q_{m',n'} \int_{-\infty}^{+\infty} e^{j\pi \beta t^2} e^{-j2\pi n \Delta f (t-\tau)} e^{j2\pi f_d t} \nonumber\\[1mm]
&& \quad \times \text{rect}\!\left(\frac{t-\tau-mT_o}{T_o}\right)
\text{rect}\!\left(\frac{t-m'T_o}{T_o}\right) \, dt.
\label{eq:i_3}
\end{IEEEeqnarray}
Similarly to \(I_2\), we can express \eqref{eq:i_3} in terms of Fresnel integrals as:
\begin{equation}
\begin{aligned}
    I_3 &= \sum_{m,n} \sum_{m',n'} X_m^*(n) \, q_{m',n'} \, 
    e^{-j2\pi n \Delta f \tau} \frac{1}{j\pi \beta} 
    e^{-j\frac{\kappa^2}{4\pi \beta}} \\
    &\;\times \big[ C(\beta u_{\max}) - C(\beta u_{\min}) 
    - j (S(\beta u_{\max}) - S(\beta u_{\min})) \big],
\end{aligned}
\end{equation}
where \(
\kappa = 2\pi f_d - 2\pi n \Delta f
\). Finally
\( I_4 \) represents the ambiguity of an OFDM signal denoted as
\begin{equation}
\begin{aligned}
   I_4&=  \sum_{m,n} \sum_{m',n'} X_{m}(n) X_{m'}^*(n') e^{j 2 \pi n' \Delta f \tau} \\ &\quad \times 
    \int_{-\infty}^{+\infty} e^{j 2 \pi \left[ (n - n') \Delta f + f_d \right] t} \\
    & \times\text{rect} \left( \frac{t - m T_{\text{o}}}{T_{\text{o}}} \right) 
    \text{rect} \left( \frac{t - \tau - m' T_{\text{o}}}{T_{\text{o}}} \right) dt.
    \label{eq:ambiguity_ofdm_formation}
\end{aligned}
\end{equation}
\vspace{-0.6em}
The product of the rectangular windows restricts \(t\) to the overlap and based on \eqref{eq:tmin_max} and \eqref{eq:ta_d}, we can write \(I_4\) as
\begin{equation}
\begin{aligned}
    I_4&= \sum_{m,n} \sum_{m',n'} X_{m}(n) X_{m'}^*(n') e^{j 2 \pi n' \Delta f \tau} \\
    &\quad \times 2 T_d \text{sinc} \left( 2 \pi \left[ (n - n') \Delta f + f_d \right] T_d \right) \\
    &\quad \times e^{j 2 \pi \left[ (n - n') \Delta f + f_d \right] T_a}.
    \label{eq:ambiguity_ofdm}
\end{aligned}
\end{equation}
This function is defined for two cases depending on the value of \(\tau\) defined in \eqref{eq:Td} and  \eqref{eq:Ta}. {The transmitted AAC-OFDM waveform is
\(
a(t) = (1-\alpha)s(t) + \alpha c(t).
\)
Substituting this into the ambiguity function yields
\(
\chi(\tau,f_d)
= (1-\alpha)^2 I_4(\tau,f_d)
+ \alpha(1-\alpha)\big(I_2(\tau,f_d) + I_3(\tau,f_d)\big)
+ \alpha^2 I_1(\tau,f_d),
\)
so that $\alpha$ only changes the relative weights of $I_1, I_2, I_3$ and  $I_4$ without altering their qualitative shapes. Allowing general chirp rates $\beta_{m,n}$ across symbols and subcarriers affects only the internal phase and the delay–Doppler coupling inside these terms, while the four-term structure and overlap windows remain unchanged. 
}
\vspace{-0.5\baselineskip}
\subsection{Ambiguity Function Definition for CM-OFDM}
The ambiguity function for CM-OFDM is defined as:
\begin{IEEEeqnarray}{rcl}
\chi(\tau, f_d) = \int_{-\infty}^{+\infty}k(t) k^*(t - \tau) e^{j2\pi f_d t} \, dt.
\end{IEEEeqnarray}
Exploiting the expression for the CM-OFDM signal from \eqref{eq:k_t} and expanding the summations over \( m, n, m', n' \):
\begin{IEEEeqnarray}{rCl}
\chi(\tau, f_d) &=& \sum_{m,n} \sum_{m',n'} X_m(n)q_{m,n} X_{m'}^*(n')q^*_{m',n'} e^{-j2\pi n' \Delta f \tau} \nonumber\\
&& \times \int_{-\infty}^{+\infty} e^{j2\pi [(n-n') \Delta f + f_d] t} e^{j\pi \beta (t^2 - (t-\tau)^2)} \nonumber\\
&& \times \text{rect} \left(\frac{t - m T_{\text{o}}}{T_{\text{o}}}\right) 
\text{rect} \left(\frac{t - \tau - m' T_{\text{o}}}{T_{\text{o}}}\right) \, dt.
\label{eq:cm_ofmd_ambiguity_eq}
\end{IEEEeqnarray}
The product of the rectangular windows restricts \(t\) to the overlap while using \eqref{eq:tmin_max} and \eqref{eq:ta_d}.
The final ambiguity function for CM-OFDM is defined as:
\begin{IEEEeqnarray}{rCl}
\chi(\tau, f_d) &=& \sum_{m,n} \sum_{m',n'} X_m(n)q_{m,n} X_{m'}^*(n')q^*_{m',n'} e^{-j2\pi n' \Delta f \tau} \nonumber\\
&& \times 2 T_d \, \text{sinc} \left(2 \pi \kappa T_d \right)  e^{j 2\pi \kappa T_a} e^{j\pi \beta (2\tau T_a - \tau^2)},
\label{eq:chi_tau_fd}
\end{IEEEeqnarray}
where \(\kappa = (n - n') \Delta f + f_d - \beta \tau.\)
\begin{comment}
\begin{IEEEeqnarray}{rcl}
\kappa = (n - n') \Delta f + f_d - \beta \tau.
\end{IEEEeqnarray}
\end{comment}
This function is defined for two cases depending on the value
of \(\tau\) considered in \eqref{eq:Td} and  \eqref{eq:Ta}.
\vspace{-0.6em}
\subsection{PAPR of AAC--OFDM }
This section quantifies how the AAC-OFDM weight 
$\alpha$ influences the peak-to-average power ratio (PAPR) of the transmitted symbol. 
We derive a bound that yields a simple threshold on $\alpha$, ensuring that the PAPR of 
AAC-OFDM is strictly lower than that of OFDM under equal-power normalization.
For any length-$N$ block $x(l)$, we define the PAPR as
\begin{equation}
\mathrm{PAPR}(x)\ \triangleq\
\frac{\displaystyle \max_{0\le l\le N-1} |x(l)|^2}{
\displaystyle \frac{1}{N}\sum_{l=0}^{N-1} |x(l)|^2 }.
\label{eq:PAPR-def}
\end{equation}
From \eqref{eq:ofdm_signal_form} \(s(l)\) denotes the OFDM time block and from \eqref{eq:chirp_signal_form} \(c(l)\) the chirp on the same useful \(N\) samples. Both are normalized to unit average power: \(\frac{1}{N}\sum_{l=0}^{N-1}\lvert s(l)\rvert^{2}=1\) and \(\frac{1}{N}\sum_{l=0}^{N-1}\lvert c(l)\rvert^{2}=1\).
%\end{comment}
Let us define the OFDM peak as \(g \triangleq \max_{0 \le l < N-1} \lvert s(l) \rvert\). Then \(\mathrm{PAPR}(s)=g^{2}\).
and for a unit-modulus chirp ($|c(l)|=1$) we have $\mathrm{PAPR}(c)=1$. The AAC-OFDM from \eqref{eq:aac-ofdm}
\(a(l)=(1-\alpha)s(l)+\alpha c(l)\), with \(\alpha\in[0,1]\), The per-symbol correlation and its magnitude are \(\rho \triangleq \frac{1}{N}\sum_{l=0}^{N-1} s(l)\,c^{*}(l)\) and \(e \triangleq |\rho|\).
The average power of $a(l)$ can be written as
\begin{equation}
G(\alpha)\ =\ (1-\alpha)^2+\alpha^2+2\alpha(1-\alpha)\,\Re\{\rho\}.
\label{eq:G}
\end{equation}
For a fair PAPR comparison at equal power, we use
\begin{equation}
u_\alpha(l) \ \triangleq\ \frac{a(l)}{\sqrt{G(\alpha)}}.
\label{eq:u}
\end{equation}
Hence, $u_0(l)=s(l)$ with $\mathrm{PAPR}(u_0)=g^2$, and $u_1(l)=c(l)$ with $\mathrm{PAPR}(u_1)=1$.
We first upper-bound the instantaneous peak via the triangle inequality and lower-bound the average power using the block correlation. Combining these bounds yields an explicit bound on $\mathrm{PAPR}(u_\alpha)$ and a corresponding design threshold on $\alpha$.
Since both $\max_l|a(l)|$ and $G(\alpha)$ vary continuously with $\alpha$, the mapping $\alpha\mapsto \mathrm{PAPR}(u_\alpha)$ is continuous within $[0,1]$ and moves from $g^2$ at $\alpha=0$ to $1$ at $\alpha=1$. Thus some $\alpha^\star\in(0,1]$ satisfies $\mathrm{PAPR}(u_{\alpha^\star})<g^2$, and $\mathrm{PAPR}(u_\alpha)\to 1$ as $\alpha\to 1$.

\textbf{Deterministic upper bound and threshold:}
From the triangle inequality, we have $\max_l|a(l)|\le (1-\alpha)g+\alpha$. Using $\Re\{\rho\}\ge -|\rho|=-e$ we obtain the lower bound $G(\alpha)\ge (1-\alpha)^2+\alpha^2-2\alpha(1-\alpha)e$. Combining with \eqref{eq:PAPR-def} gives
\begin{equation}
\mathrm{PAPR}(u_\alpha)\ \le\
\frac{\big((1-\alpha)g+\alpha\big)^2}{
(1-\alpha)^2+\alpha^2-2\alpha(1-\alpha)\,e}.
\label{eq:papr-bound}
\end{equation}
A sufficient condition for \(\mathrm{PAPR}(u_\alpha)<g^{2}\) is that the right-hand side of (47) is lower than \(g^{2}\); enforcing this inequality gives the threshold
\begin{equation}
\alpha_{\mathrm{th}}(g,e)=\frac{2g\,(ge+1)}{g^{2}+2g-1+2g^{2}e},
\label{eq:alpha-th}
\end{equation} 
and therefore any \(\alpha>\alpha_{\mathrm{th}}(g,e)\) ensures \(\mathrm{PAPR}(u_\alpha)<g^{2}\).
%\vspace{-0.9em}
As a design implication, using \eqref{eq:papr-bound}, any choice \(\alpha>\alpha_{\mathrm{th}}(g,e)\) in (2) guarantees \(\mathrm{PAPR}(u_\alpha)<\mathrm{PAPR}(s)=g^{2}\) under equal-power normalization; in the decorrelated regime (\(e\approx 0\)), the threshold reduces to \(\alpha_{\mathrm{th}}(g,0)=\dfrac{2g}{g^{2}+2g-1}\).

\vspace{-1em}
 \subsection{Design of Sensing receiver for AAC-OFDM and CM-OFDM}
At the receiver side, the range estimation is performed with a matched filter over a finite observation window. The CPI begins at $t_0$ and has a duration of $T_{\mathrm{obs}}$. In symbol-wise processing $T_{\mathrm{obs}}=T_o$, and in slot-wise processing with $M$ symbols $T_{\mathrm{obs}}=M T_o$. The window
\(
W(t)=\operatorname{rect}\!\left(\frac{t-t_0}{T_{\mathrm{obs}}}\right)
\)
restricts processing to $[t_0,\,t_0+T_{\mathrm{obs}})$. The physical round-trip delay $\tau_0=2R/c$ is embedded in the received waveforms, and the matched filter sweeps a lag variable $\tau$ to locate the peak.

Using $r(t)$ from (10) and the chirp template $c(t)$ from (23), the AAC-OFDM matched-filter output over the observation window is
\begin{equation}
\Upsilon_{\mathrm{AAC}}(\tau)
= \int_{-\infty}^{\infty} r(t)\,c^{*}\!\big(t-\tau\big)\,W(t)\,dt .
\label{eq:aac-mf}
\end{equation}
This evaluation requires only the chirp rate $\beta$ and no data symbols are needed.

Using $y(t)$ from (13) and the CM-OFDM waveform $k(t)$ from (12), the CM-OFDM matched-filter output is
\begin{equation}
\Upsilon_{\mathrm{CM}}(\tau)
= \int_{-\infty}^{\infty} y(t)\,k^{*}\!\big(t-\tau\big)\,W(t)\,dt ,
\label{eq:cm-mf}
\end{equation}
which requires the knowledge of $\beta$ and the data sequence $X_m(n)$ through $k(t)$.
\begin{comment}
    
The delay estimate is obtained by maximizing the magnitude of the matched-filter output for the waveform under test as follows:
\begin{equation}
\hat{\tau}_{\chi}=\arg\max_{\tau}\, \big|\Upsilon_{\chi}(\tau)\big|, 
\quad \chi\in\{\text{AAC-OFDM},\,\text{CM-OFDM}\},
\tag{57}\label{eq:tau-est}
\end{equation}
and the corresponding range estimate is
\begin{equation}
\hat{R}_{\chi}=\frac{c}{2}\,\hat{\tau}_{\chi}, 
\qquad \chi\in\{\text{AAC-OFDM},\,\text{CM-OFDM}\}.
\tag{58}\label{eq:range-est}
\end{equation}
Velocity follows from the Doppler-induced phase progression across OFDM symbols within the CPI: after dechirping, unwrap the per-symbol peak phases of $\Upsilon_{\chi}(\tau)$ and fit a linear slope to obtain $\hat{f}_{d,\chi}$, yielding $\hat{v}_{\chi}=\tfrac{\lambda}{2}\,\hat{f}_{d,\chi}$ in accordance with the received-signal models in (10) and (13).
\end{comment}
In practice \cite{SmithFFTConv}, \eqref{eq:aac-mf} and \eqref{eq:cm-mf} are evaluated via FFT-based circular correlation per OFDM symbol: one $N$-point FFT of the received block, a pointwise multiply with the precomputed spectrum of the template, and one $N$-point IFFT, while for AAC-OFDM the template spectrum is fixed within a CPI, whereas for CM-OFDM it is regenerated each symbol.
\vspace{-0.6em}
\subsection{Computational Complexity Analysis}
We compare the receiver-side complexity of the proposed AAC--OFDM and CM--OFDM sensing pipelines against conventional OFDM sensing based on correlation with a fixed reference such as PRS. Throughout, we adopt the standard radix-2 FFT cost model, where an $N$-point FFT or IFFT requires $(N/2)\log_{2}N$ complex multiplications.

For AAC--OFDM sensing, each symbol in a coherent processing interval (CPI) is processed using FFT-based circular correlation. The received block undergoes an $N$-point FFT, and then it is multiplied pointwise with the precomputed spectrum of the chirp template, followed by an $N$-point IFFT. This results in $N\log_{2}N + N$ complex multiplications per symbol. Since the chirp is data-independent, its spectrum is computed once per CPI at a cost of $(N/2)\log_{2}N$. The overall sensing complexity of AAC--OFDM over $M$ symbols is
\begin{equation}
C^{\text{AAC-OFDM}}_{\text{sense}} = M\!\left(N\log_{2}N + N\right) + \frac{N}{2}\log_{2}N .\label{eq:comp_aac}
\end{equation}

For CM--OFDM sensing, the matched filter template depends on the transmitted data and must be regenerated for each symbol. Every symbol therefore requires two $N$-point FFTs, one pointwise multiplication, and one $N$-point IFFT, which amounts to $1.5N\log_{2}N + N$ complex multiplications per symbol. The resulting complexity over the CPI is
\begin{equation}
C^{\text{CM-OFDM}}_{\text{sense}} = M\!\left(1.5N\log_{2}N + N\right).
\label{eq:comp_cm}
\end{equation}
When CM-OFDM is used to enhance PRS as shown in Fig.~4, the reference is the fixed PRS multiplied by the fixed chirp, so its spectrum can be computed once and reused within a CPI. Consequently, the sensing complexity equals that of AAC-OFDM.

For conventional OFDM sensing using a fixed pilot or PRS, the reference template is constant across symbols and its spectrum can be computed once per CPI. Each symbol then requires one FFT, one IFFT, and a pointwise multiplication, which is identical to AAC--OFDM. The sensing complexity is therefore
\begin{equation}
C^{\text{OFDM}}_{\text{sense}} = M\!\left(N\log_{2}N + N\right) + \frac{N}{2}\log_{2}N .
\label{eq:comp_ofdm}
\end{equation}

On the communications path, all three schemes require one FFT per symbol for OFDM demodulation followed by per-subcarrier equalization. AAC--OFDM and CM--OFDM add only a dechirp operation in the time domain, which amounts to $N$ complex multiplications per symbol and it is negligible compared with the FFT cost.

Finally, Doppler estimation implemented through phase unwrapping across the $M$ slow-time samples, differencing, and averaging involves $\mathcal{O}(M)$ lightweight scalar operations, which is negligible compared to the $\mathcal{O}(MN\log N)$ FFT workload that dominates the receiver complexity. AAC--OFDM achieves the same asymptotic complexity as conventional OFDM sensing while operating without pilots, and it is more efficient than CM--OFDM by eliminating the per-symbol template FFT.
\section{Proposed Implementation Technique to Improve Sensing}\label{sec:slot_vs_symbol}
\begin{figure*}[!htb]
\centering
\subfloat[]{\includegraphics[width=0.35\linewidth,  trim=0pt 0pt 0pt 0pt,clip]{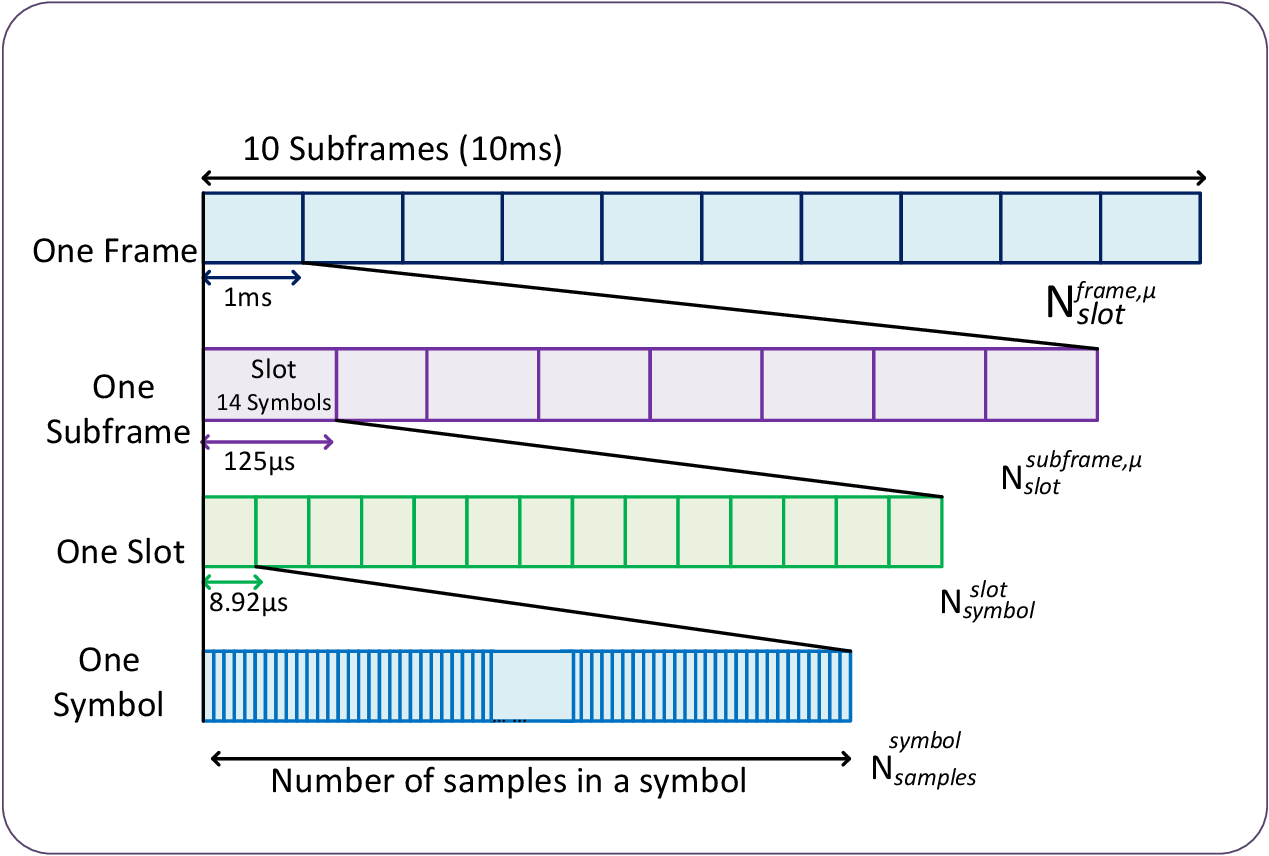}}\label{fig:block1_time_slot}
\subfloat[]{\includegraphics[width=0.24\linewidth,trim=0pt 0pt 0pt 0pt,clip]{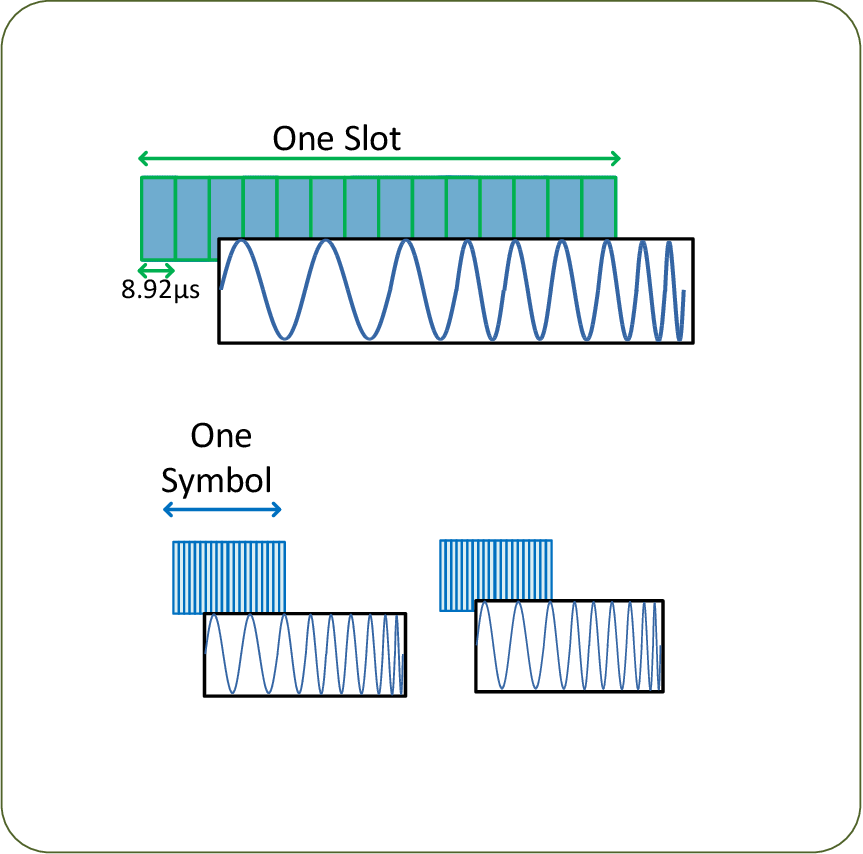}\label{fig:block2_time_slot}}
\subfloat[]{\includegraphics[width=0.245\linewidth,  trim=0pt 0pt 0pt 0pt,clip]{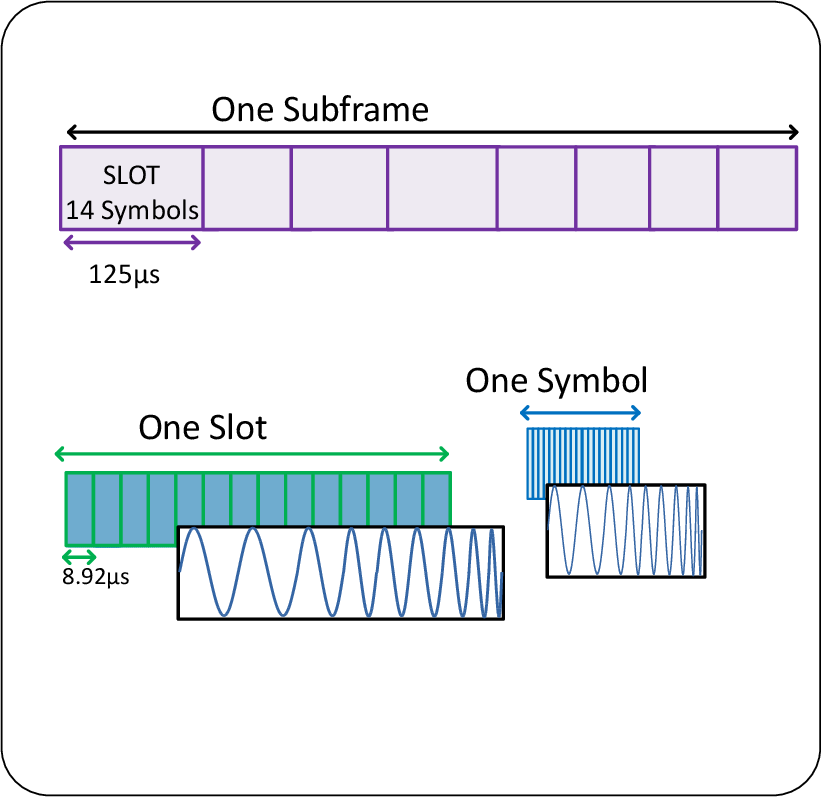}\label{fig:block3_time_slot}}
\caption{5G NR Frame structure. (a) Frame structure along with the demonstration of each sample in the symbol. (b) Slot-wise and symbol-wise chirp incorporation. (c) Hybrid incorporation of chirp in time domain for one subframe.}\label{fig:5gnr_time}
\end{figure*}
In this section, we propose a technique for improving the range and velocity estimation by developing a novel implementation technique based on the 5G New Radio (NR) frame structure.\vspace{-0.75em}
\subsection{Slot and Symbol-Based Implementation}\label{sec:slot_vs_symbol1}
 In Fig. \ref{fig:5gnr_time}(a), we show the 5G frame structure where a frame of duration 10ms has 10 subframes of 1ms duration each. In this paper, we consider the subcarrier spacing \(\Delta f\) to be 120kHz and each subframe has 8 slots of 125\(\mu s\)\footnote{{120kHz} subcarrier spacing is used for illustration in this paper; the methodology is applicable to other configurations as well.}. 
 Each slot is occupied by 14 OFDM symbols \cite{3gpp_38_211}. After performing \(N\)-point IFFT on the subcarrier data and adding a cyclic prefix, the number of sampling points becomes \(N+N_{CP}\) in $T_o$ duration. In  Fig.~\ref{fig:5gnr_time}(b), we present two ways of incorporating the chirp: a slot-based and a symbol-based one. Introducing a single chirp for the whole slot leads to an increase in the maximum unambiguous range, as the duration of the chirp will be $14T_{o}$, corresponding to 14 OFDM symbols. The slot-based implementation will also improve the velocity estimation, as the Cramér–Rao lower bound for Doppler estimation shows
that increasing the observation duration reduces the velocity estimation error and thus directly improves the Doppler resolution \cite{9921271}. Our results show that introducing the chirp in the time slot significantly improves the SNR and hence also improves the RMSE of range estimation. The symbol-based implementation increases the maximum unambiguous velocity estimation. The hybrid implementation is a combination of the slot-level and symbol-level implementations within the same subframe. As illustrated in Fig.~3(c), in an eight-slot subframe we apply the chirp across the entire first slot following the slot-level mode to increase the unambiguous range, and in the second slot we confine the chirp to a single OFDM symbol following the symbol-level mode to widen the unambiguous velocity estimation window for fast moving targets. The remaining slots can either repeat this pair in the order of full slot followed by single symbol across the subframe or adapt this pattern as needed, such as a full slot in slots 0 and 4 and a single symbol in slots 1 and 5, while the OFDM frame structure and the scheduled data subcarriers remain unchanged.
 \begin{figure*}[!htb]
\centering
{\includegraphics[width=0.8\linewidth,height=4.0cm,  trim=0pt 0pt 0pt 0pt,clip]{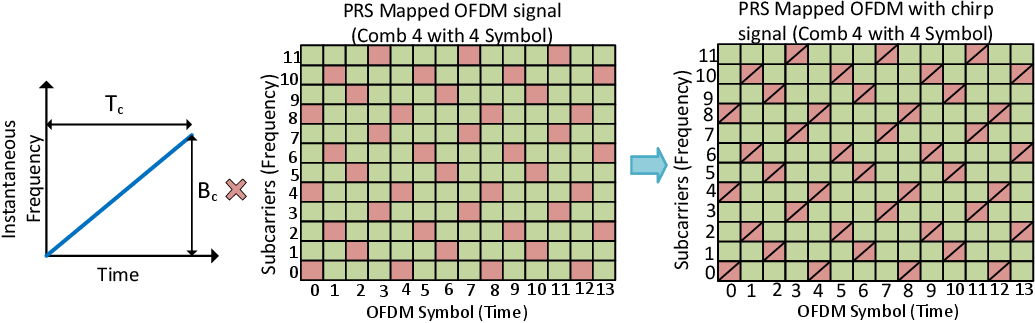}}\label{fig:frequency_time_block}
\caption{Incorporation of the CM-OFDM waveform in the OFDM time-frequency resource block to improve the PRS performance.}\label{fig:PRS_CHIP}
\end{figure*}
\vspace{-0.8em}
 \subsection{CM-OFDM enhancing PRS performance}
Fig.~\ref{fig:PRS_CHIP} shows a single 5G NR resource block (RB), comprising 12 consecutive subcarriers over 14 OFDM symbols with each cell representing a resource element (RE). Resource blocks (RBs) are aggregated to form the resource grid (RG), which is the complete two-dimensional array of all resource elements over frequency and time, as detailed in \cite{3gpp_38_211}. The CM-OFDM scheme can be incorporated into the existing time-frequency RBs of conventional OFDM systems, for enhancing the PRS performance through improved range resolution. This will be validated by the ambiguity function analysis presented later in Section \ref{sec:ambiguity_function}. Fig.~\ref{fig:PRS_CHIP} presents an example of a PRS comb 4 based mapping comprising four symbols \cite{9921271}, illustrating how a chirp is applied to the OFDM symbol that serves as the PRS symbol within the existing time–frequency structure.   Since our CM-OFDM design equation \eqref{eq:k_m} is capable of assigning this waveform to selected OFDM resource elements, we can ensure that it consistently occupies the same position, where the PRS is located across the grid.
\vspace{-0.2em}
 \subsection{Flexibility offered by AAC-OFDM}
\begin{figure*}[!htb]
\centering
{\includegraphics[width=0.8\linewidth,height=3.9cm,  trim=0pt 0pt 0pt 0pt,clip]{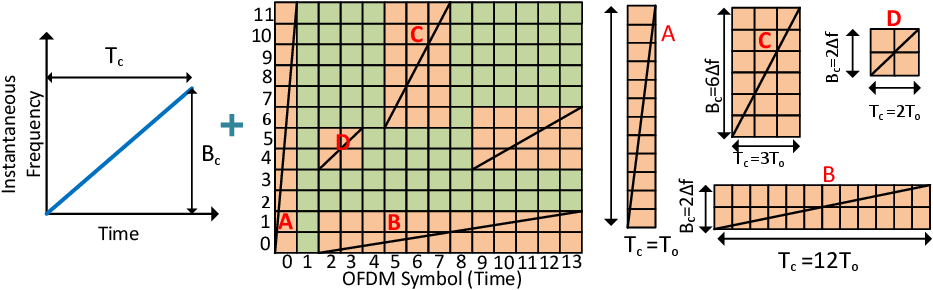}}
\caption{Flexibility of the AAC-OFDM implementation in OFDM time-frequency resources to improve the sensing task performance.}\label{fig:AAC_frequecny_time}
\end{figure*}

 Some applications require high range resolution, while others benefit from maximum unambiguous velocity. AAC-OFDM offers an enhanced solution because it can be integrated into OFDM RBs in various configurations. Fig.~\ref{fig:AAC_frequecny_time} shows the flexibility of the AAC-OFDM, which can be exploited for improving the sensing performance. In Fig. \ref{fig:AAC_frequecny_time}, Block A is shown for a single RB, and it can be extended for the entire RG to use all the frequency resources in one RG of one symbol, which provides the best range resolution and the maximum unambiguous velocity, which can be verified from Table \ref{tab:performance}. Explicitly, in Table \ref{tab:performance} we show the sensing performance metrics for the different configurations shown in Fig.~\ref{fig:AAC_frequecny_time}. 
Similarly, Block B in Fig.~\ref{fig:AAC_frequecny_time} is particularly suitable for scenarios in which maximizing the unambiguous range is the primary design requirement, where increasing the time duration of the chirp increases the maximum unambiguous range. Block C in Fig.~\ref{fig:AAC_frequecny_time} allows flexible adjustment of the number of subcarriers versus the symbol duration, making it the optimal choice for customized resource allocations. Observe in Table \ref{tab:performance} that when extending the number of subcarriers and symbols according to Block C, we can achieve the best performance for all the sensing metrics. Block D adopts square time-frequency resources, and this configuration can be reduced to a single RE, as well as extended to all the symbols. Nonetheless, because the time–frequency slope (i.e., chirp rate)  remains constant, the receiver can employ a single, fixed matched-filter based dechirping architecture simply by updating the chirp’s start and end frequencies and the corresponding time window to accommodate different time–frequency configurations.
\begin{comment}
\begin{table}[!t]
\centering
\caption{Combined Performance Metrics (with \(N=256\) fixed)}
\label{tab:performance}
\begin{tabular}{|l|c|c|c|c|c|}
\hline
\textbf{Configuration} 
  & \textbf{\(n\)} 
  & \textbf{\(m\)} 
  & \(\Delta R\) (m) 
  & \(R_{\max}\) (m) 
  & \(\lvert v\rvert\) (m/s) \\
\hline
\shortstack[l]{Max unambiguous\\Range (B)} 
  & 2   & 14  & 625.0   & \(1.20\times10^{6}\)   & 25.0    \\ \hline
High Velocity       
  & 64  & 1   & 19.53   & 2,676                & 350.6   \\ \hline
\shortstack[l]{Best Range\\Resolution (A)}    
  & 256 & 1   & 4.88    & 669                  & 350.6   \\ \hline
Optimal Value  
  & 64  & 7   & 19.53   & 18,732               & 50.1    \\ \hline
\shortstack[l]{Resolution\\and Range}  
  & 256 & 14  & 4.88    & 9,366                & 25.0    \\ \hline
\end{tabular}
\end{table}
\end{comment}

% --- compact table with local spacing tweaks ---
\begingroup
% Tighten float/caption spacing just for this table
\setlength{\textfloatsep}{6pt}
\setlength{\floatsep}{6pt}
\setlength{\abovecaptionskip}{2pt}
\setlength{\belowcaptionskip}{4pt}

\begin{table}[!t]
\centering
\caption{Combined Performance Metrics for \(N=256\)}
\label{tab:performance}

\small                 % smaller font
\setlength{\tabcolsep}{3.5pt}  % narrower columns
\renewcommand{\arraystretch}{0.9} % tighter rows

\begin{tabular}{|l|c|c|c|c|c|}
\hline
\textbf{Config.} & \textbf{\(n\)} & \textbf{\(m\)} & \(\Delta R\) (m) & \(R_{\max}\) (m) & \(|v_{\max}|\) (m/s) \\
\hline
Block A & 256 & 1  & 4.88  & 669              & 350.6 \\
\hline
Block B & 2   & 14 & 625.0 & \(1.20\times10^{6}\) & 25.0  \\
\hline
Block C & 64  & 7  & 19.53 & 18{,}732         & 50.1  \\
\hline
Block D & 2   & 2  & 625   & \(1.71\times10^{5}\) & 175.3 \\
\hline
\end{tabular}
\end{table}
\endgroup

When the chirp is introduced in a single symbol along with \(B_c=N \Delta f\) then it improves the range resolution. By contrast, when it introduces \(M\) symbols, \(T_c=M T_o\) increases the maximum unambiguous range by a factor of \(M\). Hence, the achievable range resolution, the maximum unambiguous range  $R_{\text{max}}$, and the maximum unambiguous velocity can be written as
\begin{IEEEeqnarray}{c}
    \Delta R = \frac{c}{2n \Delta f}, \quad R_{\text{max}} = \frac{c f_s mT_{o}}{4n \Delta f},\quad \lvert v_{\text{max}} \rvert < \frac{\lambda}{4 mT_{o}}.
\label{eq:range_reso_max_claim}
\end{IEEEeqnarray}
\vspace{-0.2em}

In Fig.~\ref{fig:block_diagram} we show that the receiver dechirps before OFDM demodulation, so the communications branch is that of a standard OFDM receiver. Residual  carrier frequency offset (CFO), common phase error (CPE), and
phase noise-induced inter-carrier interference are then handled as in conventional OFDM \cite{10702556,9582730}. On the sensing branch before dechirping, oscillator phase noise degrades coherent integration and raises the background noise in the range–Doppler map, making peaks wider. Hence, weak/close targets are harder to separate \cite{10102330}. This mainly reduces accuracy without changing the nominal resolutions, which are set by bandwidth and coherent processing time~\cite{7556419}. Practically, we keep measurements phase-aligned across the coherent processing interval (CPI) by estimating and removing the slowly varying common phase term of each symbol. If needed, we apply very light PN-aware smoothing, preserving the usual FFT pipeline \cite{8010432}. A small residual CFO on the sensing chain is indistinguishable from true Doppler, giving \(\hat f_d = f_d + \varepsilon\) and a velocity bias of \(\Delta v = (\lambda/2)\,\varepsilon\), while the delay over a CPI remains essentially unchanged \cite{10149610,Lee2013/07}. This can be mitigated by estimating this residual tone and derotating before Doppler estimation (or reusing the communications CFO estimate), and by maintaining Tx/Rx clock alignment to remove the bias at its source \cite{9529026}. Asynchronous impairments arising from clock drift or a small sampling-rate offset accumulate across the CPI, producing a residual delay error and range bias \(\Delta R \approx c\,\Delta t/2\), while weakening coherent integration \cite{9052489,7870764}. This can be addressed using a pilot-aided re-anchoring method \cite{s23125605}, and then coherent integration. But again, the existing FFT-based pipeline is preserved. Importantly, the quantities $\varepsilon$ and $\Delta t$ model the effective frequency 
and timing mismatch between the transmitter and receiver in both monostatic and bistatic 
operation. In the bistatic case, they capture the relative clock and sampling-rate errors 
between base stations, so the same analysis for $\Delta v$ and $\Delta R$ describes the 
impact of synchronization errors on the range--Doppler map.

\vspace{-0.4em}

\vspace{-0.5em}
\section{Simulation Results}\label{sec:simulation_results}
\begin{table}[ht]
  \centering
  \caption{Transmission and Simulation Parameters for 120\,kHz Subcarrier Spacing \cite{3gpp_38_211}}
  \label{tab:120khz}

  % ===== height tweaks (local to this table only) =====
  \small                          % or \small  / \scriptsize
  \setlength{\tabcolsep}{4pt}            % default ~6pt; smaller packs columns
  { \renewcommand{\arraystretch}{0.9}    % <1.0 shrinks row height; try 0.85–0.95
  \begin{tabular}{|l|l|}
    \hline
    \multicolumn{2}{|c|}{\bfseries Transmission Parameters} \\ \hline
    Subcarrier interval $\Delta f$ (kHz)     & 120                \\ \hline
    $N^{\mathrm{slot}}_{\mathrm{symb}}$      & 14                 \\ \hline
    $N^{\mathrm{frame}}_{\mathrm{slot}}$     & 80                 \\ \hline
    FR1 (450\,MHz--5.9\,GHz)                 & \xmark             \\ \hline
    FR2 (24.2\,GHz--52.6\,GHz)               & \checkmark         \\ \hline
    $T$ (\textmu s)                          & 8.33               \\ \hline
    $T_{\mathrm{CP}}$ (\textmu s)            & 0.57               \\ \hline
    $T + T_{\mathrm{CP}}$ (\textmu s)        & 8.92               \\ \hline
    \multicolumn{2}{|c|}{\bfseries Simulation Parameters} \\ \hline
    Carrier frequency $f_c$                  & 24\,GHz            \\ \hline
    Sampling frequency $f_s$                 & $N \times \Delta f$\\ \hline
    Modulation format                        & QPSK               \\ \hline
    Total number of subcarriers $N_c$        & 1024               \\ \hline
    Range of target $R$                      & 50\,m              \\ \hline
    Velocity of target $v$                   & 30\,m/s            \\ \hline
    Weighting factor for sensing $\alpha$    & 0.5                \\ \hline
  \end{tabular}
  }% end local arraystretch
\end{table}

%%%%%%%%%%%%%%%%%%%%%%%%%%%%%%%%%%%%%%%
\begin{table}[ht]
\centering
\caption{Channel Tap Power Levels and Delays \cite{doi:https://doi.org/10.1002/9780470825631.ch4}}
\label{tab:channel_taps_horizontal}
\setlength{\tabcolsep}{4pt}            % default ~6pt; smaller packs columns
   \renewcommand{\arraystretch}{0.9}    % <1.0 shrinks row height; try 0.85–0.95
\begin{tabular}{|l|c|c|c|c|c|}
\hline
                      \textbf{Path} & \text{ 1} & \text{2} & \text{3} & \text{ 4} & \text{5} \\
\hline
\textbf{Power (dB)}    & 0                  & -8                 & -17                & -21                & -25                \\
\hline
\textbf{Delay tap}         & 0                  & 3                  & 5                  & 6                  & 8                  \\
\hline
\end{tabular}
\end{table}
In this section, we characterize the error performance of communications and the RMSE of the range and velocity for the sensing. We assume the subcarrier spacing to be 120kHz and Table \ref{tab:120khz} represents all the parameters for the subcarrier spacing of 120 kHz used for other simulations. The delay profile for the multipath channel is given in Table
\ref{tab:channel_taps_horizontal}.
We select FR2 at 24\,GHz along with the 5G NR subcarrier spacing $\Delta f = 120\,\mathrm{kHz}$ to model urban sensing and communications in city corridors and intersections. At this carrier, a speed of $30\,\mathrm{m/s}$ induces a Doppler shift of about $4.8\,\mathrm{kHz}$, which is far lower than $120\,\mathrm{kHz}$, so the OFDM subcarrier orthogonality is preserved and Doppler is cleanly observable \cite{Feng2020FOFDM}. With $N_c = 1024$ subcarriers the occupied bandwidth is $122.88\,\mathrm{MHz}$, yielding range resolution of $\Delta R \approx 1.22\,\mathrm{m}$, when the chirp spans the full band, while keeping the sampling rate consistent with NR signal models. The channel uses a five-tap decaying power profile with path delays from $0$ to $65\,\mathrm{ns}$ to represent short-delay-spread FR2 links \cite{3GPP38901}, and the $0.57\,\mu\mathrm{s}$ cyclic prefix exceeds the approximately $65\,\mathrm{ns}$ excess delay to suppress inter-symbol interference. These settings keep the simulations standards-aligned and ensure that both range and velocity are observable under realistic urban conditions.

We utilize the classic correlation algorithm for estimating the range and velocity. First, we apply the correlation algorithm to each symbol, where the time delay can be evaluated by identifying the lag of the maximum correlation peak. Then, to determine the velocity, { we first collect the complex samples
$z_m$ for the selected range bin across the OFDM symbols in the CPI
and unwrap their phases $\phi_m = \arg(z_m)$ to eliminate discontinuities.
The differences between successive unwrapped phases,
$\Delta \phi_m = \phi_{m+1} - \phi_m$, are averaged to obtain a Doppler
frequency estimate $\hat{f}_d$, and the velocity is then quantified using
$\hat{v} = \dfrac{c}{2 f_c}\,\hat{f}_d$ \cite{10371091}.}
  For both CM-OFDM and OFDM we use the transmitted symbol as a reference to obtain the sensing information. For AAC-OFDM, the sensing information can be extracted solely from the receiver’s knowledge of the chirp rate. This characteristic offers the potential advantage of leveraging the OFDM symbol for data transmission rather than dedicating it exclusively to sensing functions, thereby making it an eminently suitable waveform for bistatic sensing scenarios.
\vspace{-1.2em}

\begin{comment}
\noindent\textit{Peak (delay) estimate:}\;
$\displaystyle \hat{\tau}\in\arg\max_{\tau}\, \bigl|y(\tau)\bigr|,\quad
y\in\{y_{\mathrm{AAC}},\,y_{\mathrm{CM}}\}.$

\paragraph{Multipath peak selection.}
In multipath, $|y(\tau)|$ exhibits several local maxima corresponding to path delays $\{\tau_\ell\}$.
We detect peaks above a threshold $\gamma$ and choose the \emph{closest} path (earliest arrival) as the LOS estimate:
\begin{align}
\mathcal{P} &\triangleq \{\;\tau\ge 0:\ |y(\tau)| \text{ is a local maximum and } |y(\tau)|\ge \gamma\;\}, \label{eq:peakset}\\
\hat{\tau}_{\text{LOS}} &\triangleq \min_{\tau\in\mathcal{P}} \tau, \qquad
\hat{R}=\tfrac{c_0}{2}\,\hat{\tau}_{\text{LOS}}. \label{eq:losselect}
\end{align}
Here $c_0$ is the speed of light. In practice we set $\gamma=\eta\,\hat{\sigma}_n$ using the off-peak noise estimate $\hat{\sigma}_n$ and a constant $\eta$ (e.g., $\eta\!\in[4,6]$), and apply a small guard interval around each detected peak to avoid picking sidelobes.
\end{comment}

\subsection{Communication Performance}

\begin{figure}
    \centering
    \includegraphics[scale=0.4]{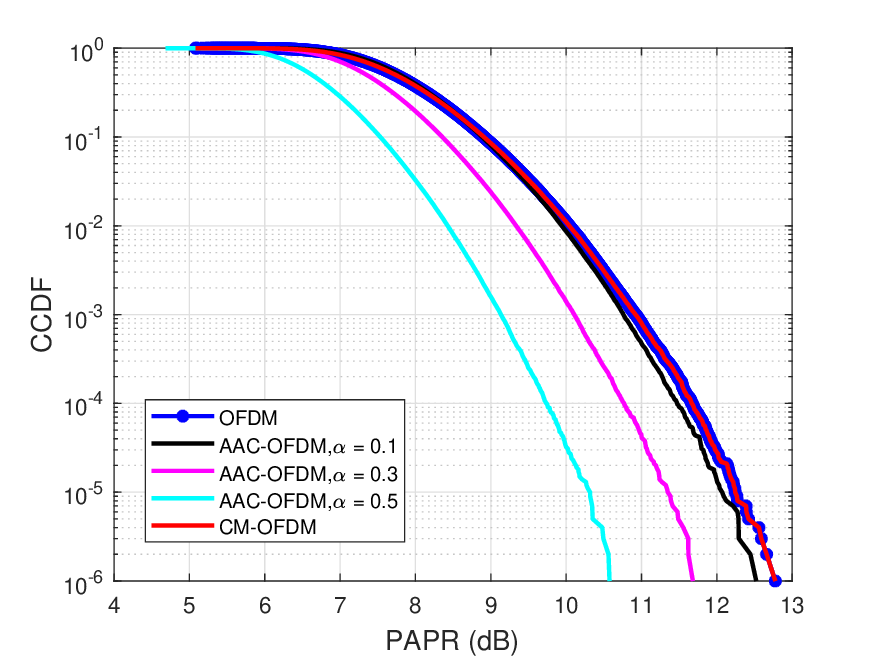} % No need to specify the full path
    \caption{CCDF of PAPR for OFDM, AAC-OFDM, and CM-OFDM.}
    \label{fig:paapr}
\end{figure}
\begin{figure}[h]
    \centering
    \includegraphics[width=0.8\linewidth,height=3.5cm,  trim=10pt 10pt 10pt 10pt,clip]{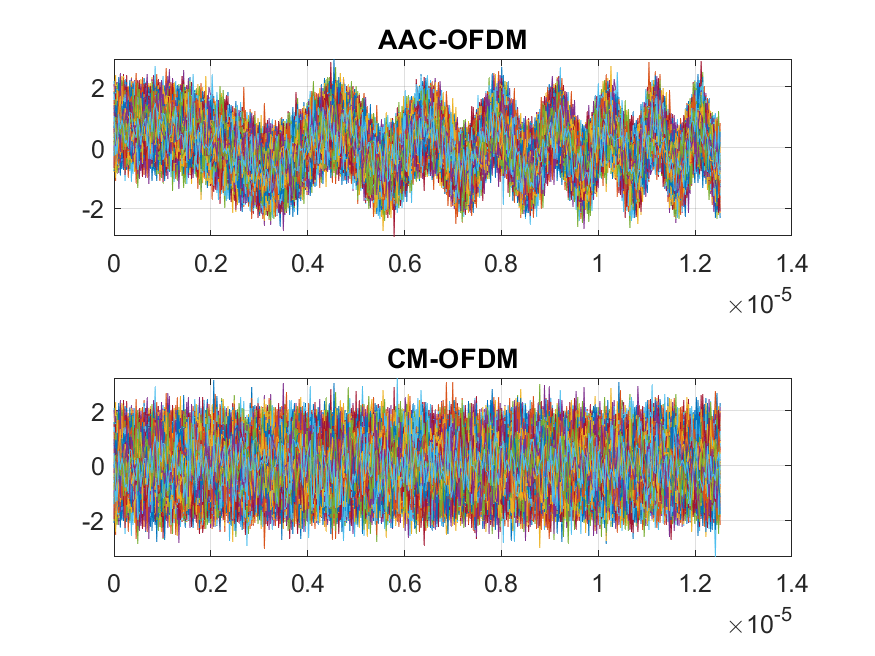} % No need to specify the full path
    \caption{Time-Domain Waveforms for AAC-OFDM and CM-OFDM.}
    \label{fig:aac_mc_waveforms}
\end{figure}\vspace{-0.7em}
\begin{figure}[h]
    \centering
\includegraphics[width=0.7\linewidth,height=4.2cm,  trim=0pt 0pt 0pt 15pt,clip]{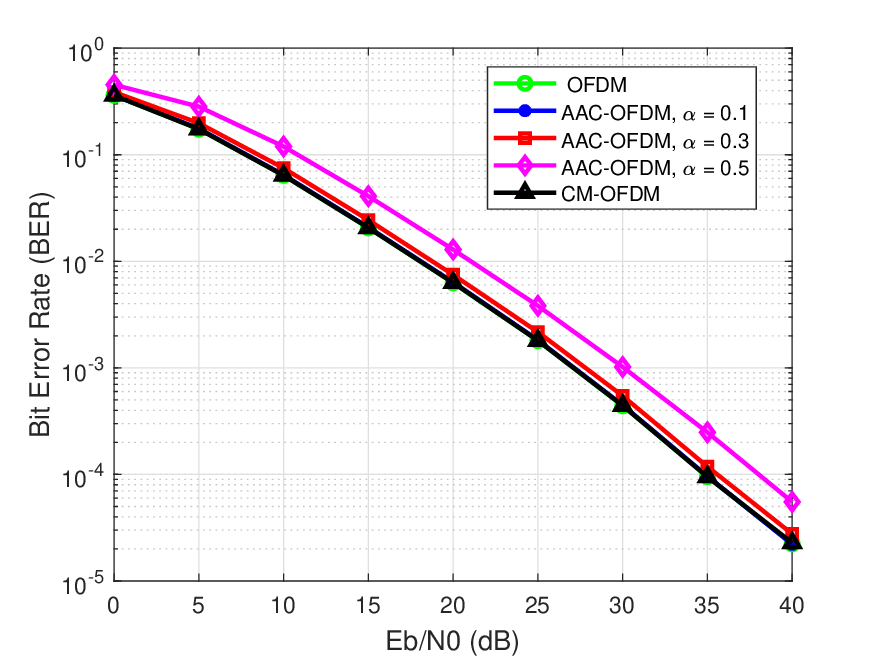} % No need to specify the full path
    \caption{BER comparison between standard OFDM, CM-OFDM and AAC-OFDM for different weighting factors $\alpha$.}
    \label{fig:BER}
\end{figure}
\vspace{0.5em}
In Fig.~\ref{fig:paapr}, we show the Complementary Cumulative Distribution Function (CCDF) of PAPR for three distinct waveforms: OFDM, AAC-OFDM and CM-OFDM. The PAPR is an essential metric in communication systems, representing the ratio of peak power to average power of a signal. The simulation parameters consist of $N=256$ for the data subcarriers, and the rest of the parameters are presented in Table~\ref{tab:120khz}.  AAC-OFDM exhibits a substantial reduction in PAPR compared to OFDM. We report quantitative values at $CCDF=10^{-3}$. At this operating point, OFDM and CM-OFDM have a PAPR of about \(11\,\mathrm{dB}\) and AAC-OFDM about \(10.9\,\mathrm{dB}\), \(10.1\,\mathrm{dB}\), and \(9\,\mathrm{dB}\) for \(\alpha=0.1,\,0.3,\) and \(0.5\), respectively.
This improvement indicates that the affine addition of the chirp function effectively smoothens the signal strength envelope, thereby reducing the likelihood of high peaks. 
 By contrast, CM-OFDM has a PAPR performance that is similar to that of standard OFDM. This suggests that the multiplicative chirp amalgam does not facilitate PAPR reduction in a manner analogous to chirp addition. The results indicate that AAC-OFDM has the potential to improve PAPR, establishing it as an appealing candidate for power-efficient communication systems.

The time-domain waveforms of AAC-OFDM and CM-OFDM are depicted in Fig.~\ref{fig:aac_mc_waveforms}, offering additional insights into their respective PAPR characteristics. For AAC-OFDM, the addition of a chirp signal produces a waveform exhibiting a smoother and more structured envelope compared to the other waveform. The introduction of periodicity through the addition of chirps mitigates abrupt fluctuations in the amplitude of the signal, thereby facilitating a PAPR reduction. The structured envelope has a more stable power level for the signal, thus mitigating significant peaks. By contrast, the waveform CM-OFDM exhibits an erratic structure, similar to that of the conventional OFDM waveform. 

The BER of the multipath channel is illustrated in Fig.~\ref{fig:BER}, where we show the results for different weighting factors of $\alpha = 0.1, 0.3, 0.5$ for AAC-OFDM. {We employ forward error correction (FEC) using a standard 1/2-rate convolutional code with generator polynomials (171, 133) in octal notation with hard-decision Viterbi decoding. This improves the BER performance under multipath and noise. Observe that AAC-OFDM behaves similarly to conventional OFDM for reduced $\alpha$ values, with $\alpha = 0.1$ achieving the minimal BER. As $\alpha$ grows to $0.3$ and $0.5$, the BER degrades since more transmit power is allocated to the chirp. CM-OFDM demonstrates comparable BER performance to conventional OFDM. These results underscore the significance of optimizing $\alpha$ for AAC-OFDM to improve the BER performance, while indicating that CM-OFDM exhibits similar performance to OFDM.}
\begin{figure}[h]
    \centering
\includegraphics[width=0.7\linewidth,height=4.2cm,  trim=0pt 0pt 0pt 18pt,clip]{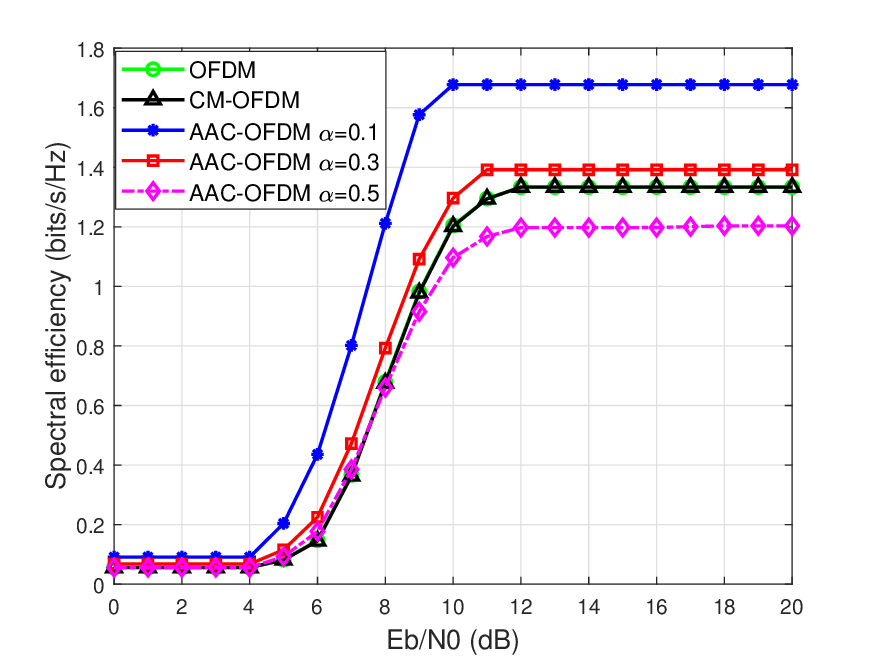} % No need to specify the full path
    \caption{Spectral efficiency comparison between standard OFDM, CM-OFDM, and AAC-OFDM for different weighting factors $\alpha$. OFDM and CM-OFDM are using 25\% of pilot subcarriers for sensing. }
    \label{fig:spec_effic}
\end{figure}
%\vspace{-0.80em}
Fig.~\ref{fig:spec_effic} shows the  spectral efficiency versus $E_b/N_0$ for OFDM, CM--OFDM, and AAC--OFDM. In OFDM and CM--OFDM, 25\% of subcarriers are reserved as pilots for sensing, which imposes a fixed throughput loss and yields plateaus at high $E_b/N_0$ of 1.3~bits/s/Hz. AAC--OFDM does not allocate pilot subcarriers, so all tones carry payload data. Consequently, AAC--OFDM achieves higher spectral efficiency across the entire $E_b/N_0$ range. Specifically, with $\alpha=0.1$ it attains 1.7~bits/s/Hz. Increasing $\alpha$ to 0.3 and 0.5 shifts the operating point to higher $E_b/N_0$ and reduces the maximum achievable spectral efficiency because more transmit power is devoted to the chirp used for sensing.
{The receiver architecture proposed in Fig.~\ref{fig:block_diagram} guarantees that the added chirp does not impair the communication link. After dechirping, the FFT demodulator operates on a standard OFDM signal, so the communication chain experiences no interference from the chirp, regardless of whether it is applied to a single symbol or to an entire slot. Thus, the choice of symbol-level versus slot-level chirp design primarily affects sensing accuracy, while communication reliability remains essentially unchanged.}

\vspace{-0.9em}
\subsection{Sensing Performance}
\begin{figure}
\centering
\subfloat[]{\includegraphics[width=0.52\linewidth]{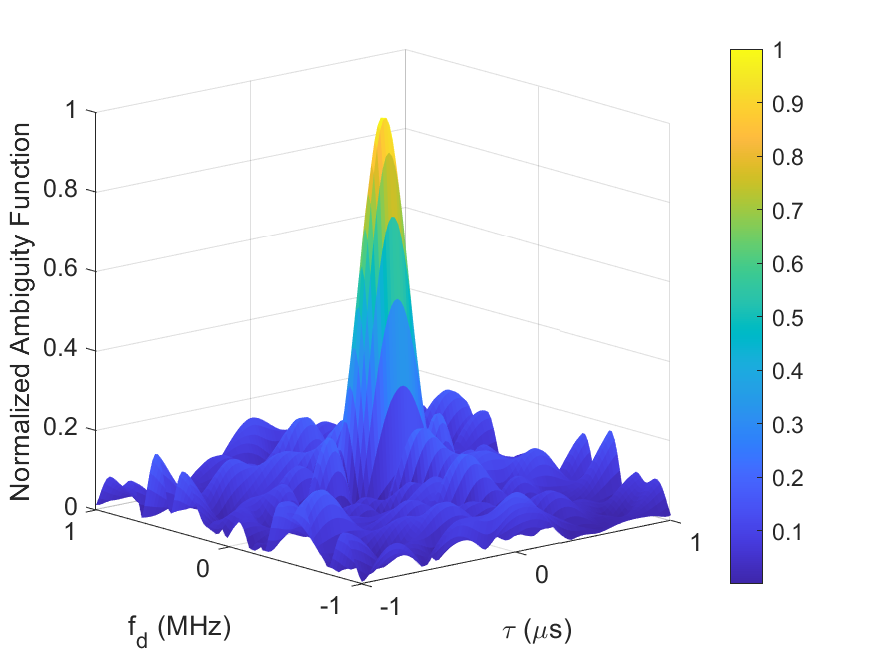}\label{fig:ofdm_only}}
\subfloat[]{\includegraphics[width=0.52\linewidth]{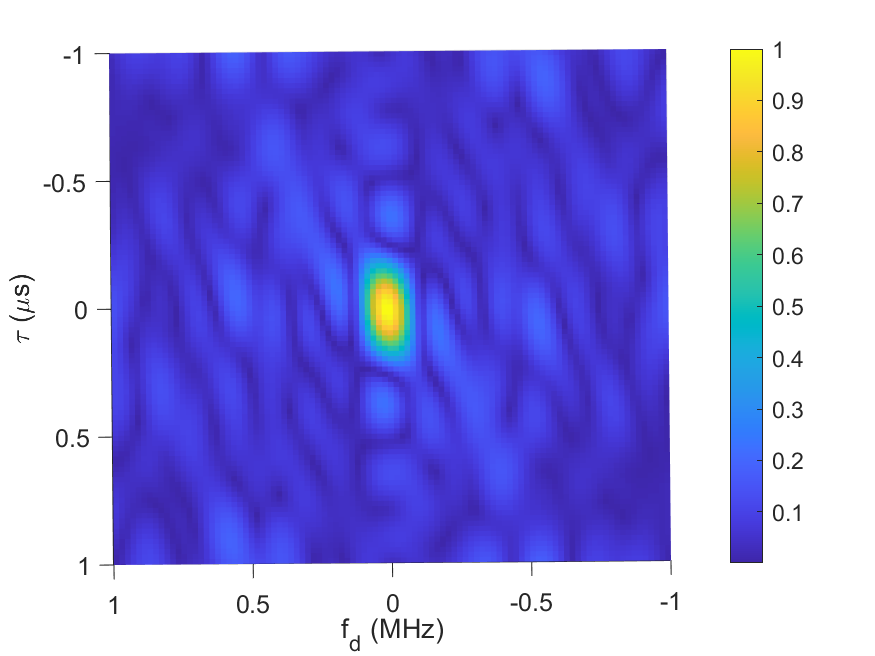}\label{fig:ofdm_only_contour}}
\caption{Ambiguity function: (a) Normalized ambiguity function of OFDM. (b) Contour view ambiguity function of OFDM.}\label{fig:ofdm_ambiguity}
\end{figure}
\begin{figure}
\centering
\subfloat[]{\includegraphics[width=0.52\linewidth]{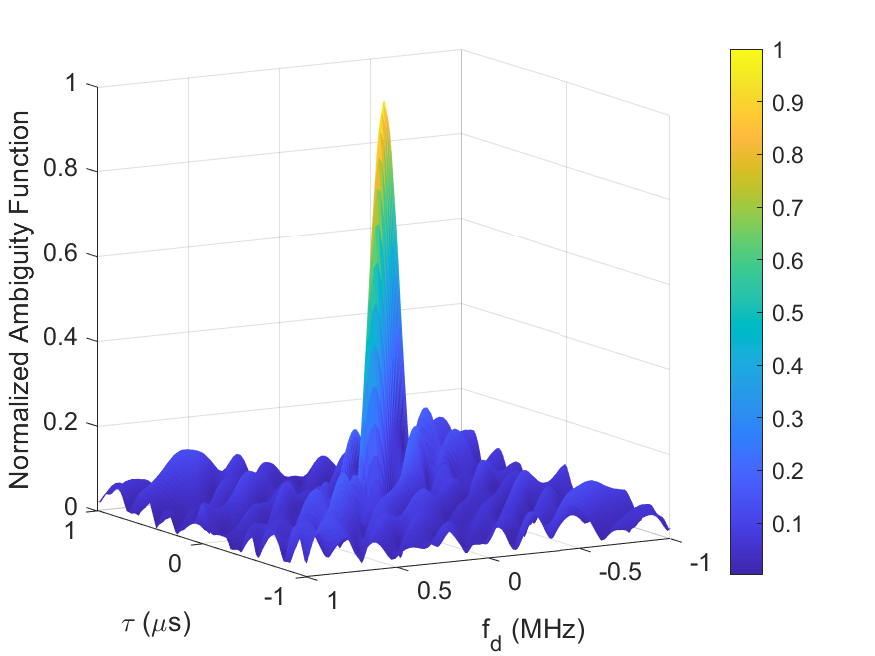}\label{fig:ofdm_mul_only}}
\subfloat[]{\includegraphics[width=0.52\linewidth]{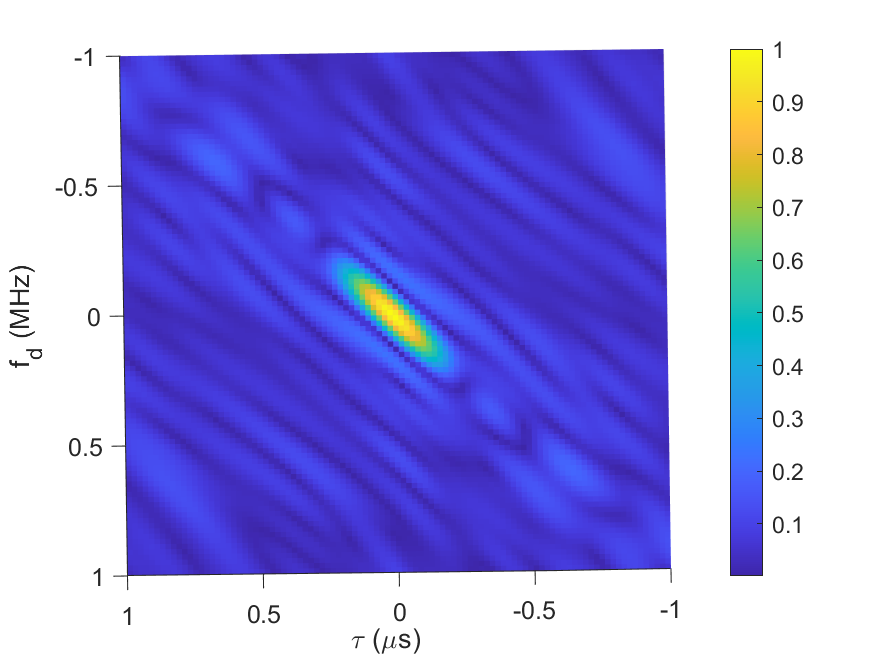}\label{fig:ofdm_mul_contour}}
\caption{Ambiguity function: (a) Normalized ambiguity function of CM-OFDM. (b) Contour of Ambiguity Function CM-OFDM.}\label{fig:ofdm_mul_ambiguity}
\end{figure}

\begin{figure}
\centering
\subfloat[]{\includegraphics[width=0.50\linewidth]{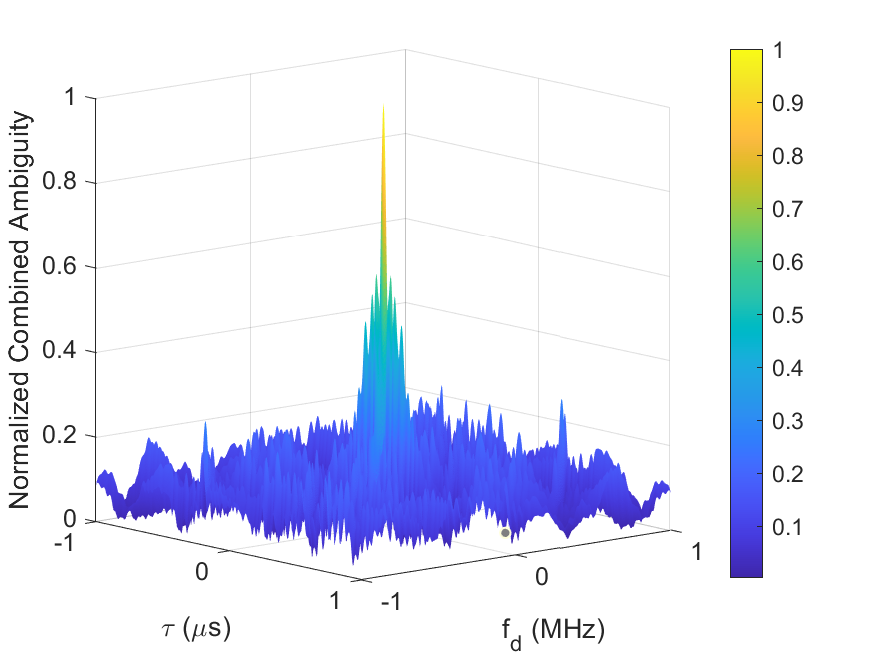}}\label{fig:acc_ofdm}
\subfloat[]{\includegraphics[width=0.48\linewidth]{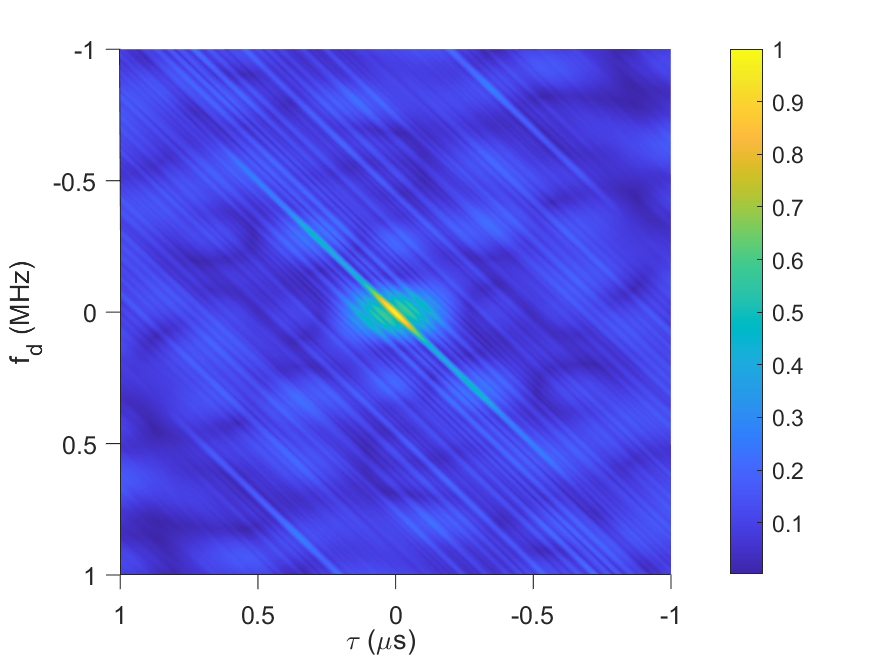}\label{fig:aac_ofdm_contour}}
\caption{Ambiguity function: (a) Normalized ambiguity function of AAC-OFDM. (b) Contour view ambiguity function of AAC-OFDM.}\label{fig:aac_ofdm_ambiguity}
\end{figure}

The ambiguity function of the OFDM waveform is shown in Fig.~\ref{fig:ofdm_ambiguity}. Fig.~\ref{fig:ofdm_ambiguity}(a) shows that it is characterized by a single sharp peak at the origin and near-zero values elsewhere. This shape indicates that the waveform succeeds in precisely distinguishing targets having different ranges and velocities. This is achieved by minimizing ambiguities in both delay and Doppler measurements. Fig.~\ref{fig:ofdm_ambiguity}(b) shows that the contour of the OFDM ambiguity function is ellipsoidal, with the vertices along the axes determining the range resolution.
With \(N=128\) and of one-symbol CPI, the delay main lobe yields \(\Delta\tau \approx 0.066~\mu\text{s}\), giving a range resolution \(\Delta R =9.89~\text{m}\) for the bandwidth employed.
We use a single symbol so the time is identical across waveforms, and the plots reflect the waveform’s intrinsic behavior. The Doppler resolution is determined by the symbol duration, and hence for all three waveforms have \(\Delta f_d  \approx 11.17\times 10^{4}\ \text{Hz}\) as shown in Fig.~\ref{fig:ofdm_ambiguity}(b), Fig.~\ref{fig:ofdm_mul_ambiguity}(b) and Fig.~\ref{fig:aac_ofdm_ambiguity}(b). For a longer CPI of \(M=128\) symbols, the Doppler resolution tightens to \(\Delta f_d \approx 11.17\times 10^{4}/128 = 872.66~\text{Hz}\), which at \(24~\text{GHz}\) gives \(\Delta v \approx (\lambda/2)\,\Delta f_d \approx 5.45~\text{m/s}\).
 The ambiguity function of CM-OFDM is plotted in Fig.~\ref{fig:ofdm_mul_ambiguity}, where Fig.~\ref{fig:ofdm_mul_ambiguity}(a) shows that the ambiguity function has a thumbtack and a tilted shape, representing both the inherited quality of OFDM and the chirp. The tilt depicted in Fig.~\ref{fig:ofdm_mul_ambiguity}(b) is due to delay-Doppler coupling. The tilt provides improved resolution in the zero-delay cut.
Since typical Doppler shifts are much smaller compared to the waveform's bandwidth, the peak values of non-zero Doppler cuts and hence the peak values of the matched filter outputs decrease only slightly when moving targets are observed. This behavior of the ambiguity function is what makes the linear FM waveform Doppler tolerant.
With \(N=128\) and a one-symbol CPI, the ambiguity plot shows a tighter delay focus than OFDM.
The measured delay main lobe gives \(\Delta\tau \approx 0.050~\mu\text{s}\), hence the range resolution is \(\Delta R \approx 7.50~\text{m}\).
 The tilt not only improves the Doppler tolerance, but also improves the delay resolution \cite{doi:https://doi.org/10.1002/0471663085.ch3}.
Fig.~\ref{fig:aac_ofdm_ambiguity} represents the ambiguity function of the AAC-OFDM. Fig.~\ref{fig:aac_ofdm_ambiguity}(a) shows the sharper thumbtack model along with tilt, which improves the range resolution along with the Doppler tolerance. In Fig.~\ref{fig:aac_ofdm_ambiguity}(b) the narrow contour represents a better range resolution for the proposed waveform.
With \(N=128\) and a one-symbol CPI, the ambiguity plot shows the sharpest delay focus among the three waveforms.
The measured delay main lobe gives \(\Delta\tau \approx 0.028~\mu\text{s}\), hence the range resolution is \(\Delta R \approx 4.20~\text{m}\).

\begin{comment}
\begin{figure}
    \centering
    \includegraphics[scale=0.40]{Pictures/mywork/new_rmse_fig_1_symbol.eps} % No need to specify the full path
    \caption{Range RMSE versus SNR when chirp is introduced for one symbol.}
    \label{fig:cor_rmse_symbol}
\end{figure}
\begin{figure}
    \centering
    \includegraphics[scale=0.40]{Pictures/mywork/new_slot_rmse_range.eps} % No need to specify the full path
    \caption{Range RMSE versus SNR when chirp is introduced for one entire slot.}
    \label{fig:cor_rmse_slot}
\end{figure}
\begin{figure}
    \centering
    \includegraphics[scale=0.40]{Pictures/mywork/pilot_subcarrier.eps} % No need to specify the full path
    \caption{Range RMSE versus SNR when chirp is introduced for one entire slot.}
    \label{fig:piot_subcarrier}
\end{figure}
\end{comment}
\begin{figure*}
\centering
\subfloat[]{\includegraphics[width=0.3\linewidth]{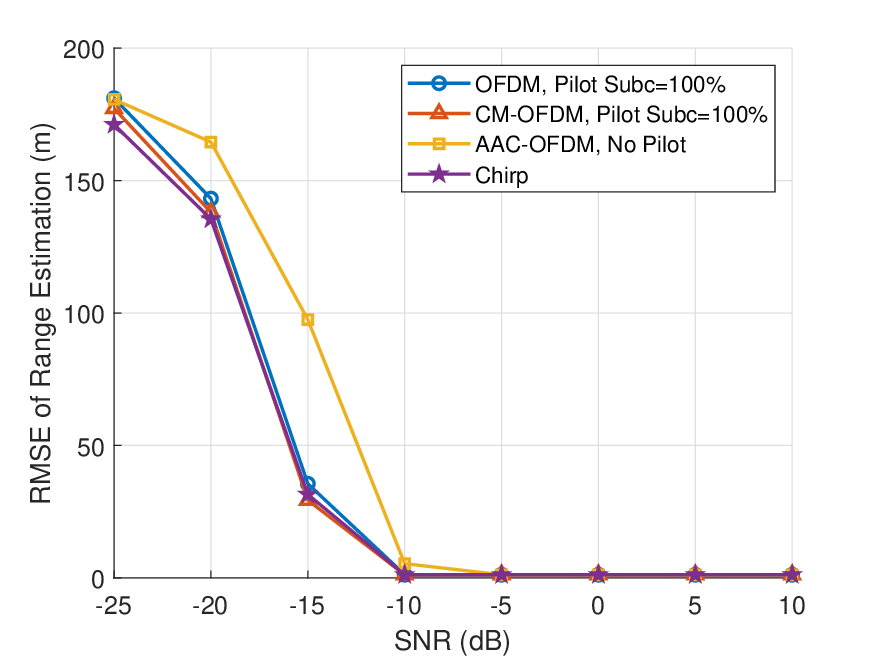}\label{fig:cor_rmse_symbol}}
\subfloat[]{\includegraphics[width=0.3\linewidth]{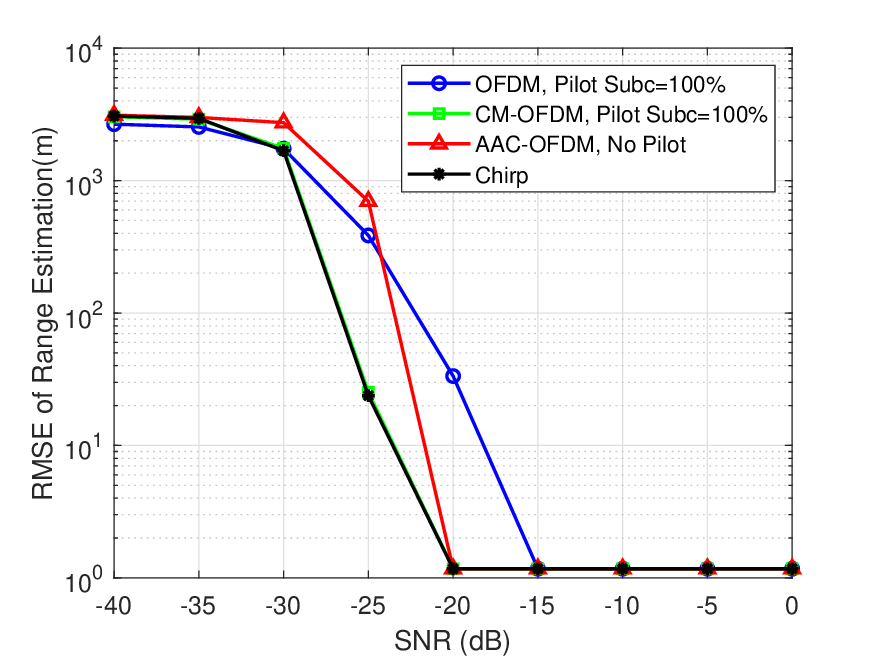}\label{fig:cor_rmse_slot}}
\subfloat[]{\includegraphics[width=0.3\linewidth]{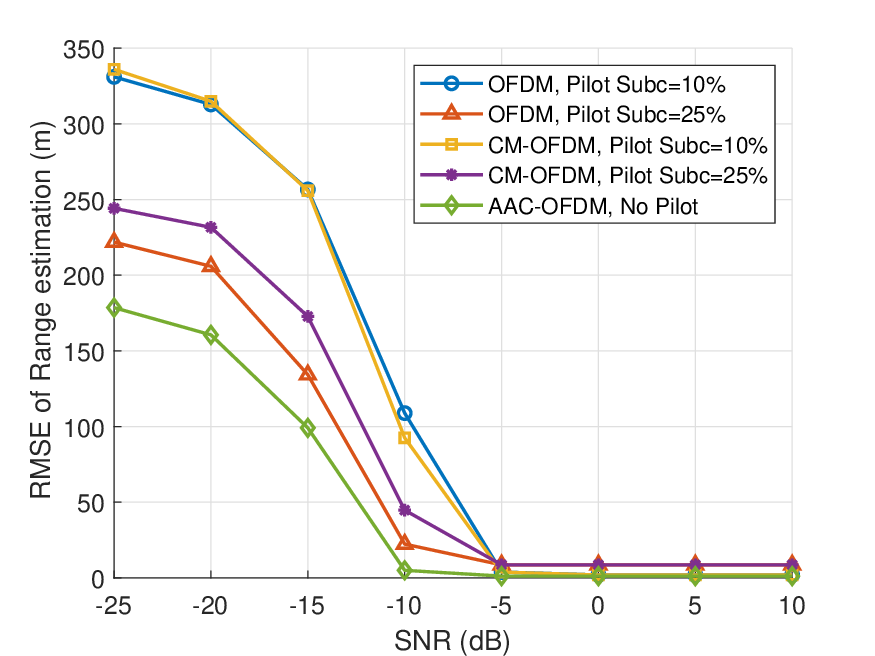}\label{fig:piot_subcarrier}}
\caption{Range RMSE versus SNR. (a) Range RMSE when chirp is introduced for one symbol. (b) Range RMSE when chirp is introduced for one entire slot. (c) Comparison of range RMSE for 10\% and 25\% of pilot subcarriers used for sensing in OFDM and CM-OFDM.}\label{fig:range_rmse}
\end{figure*}

Fig.~\ref{fig:range_rmse}(a) and Fig.~\ref{fig:range_rmse}(b) show the RMSE performance for different waveforms while considering two distinct cases: the incorporation of the chirp signal within a single OFDM symbol and its incorporation across an entire slot. In both sets of results, we assume that CM-OFDM and OFDM are using all the subcarriers as a pilot to carry out sensing. Fig.~\ref{fig:range_rmse}(a)  focuses on the RMSE performance when the chirp is introduced for a single OFDM symbol. At low SNRs (e.g., -30 to -15 dB), all waveforms have significantly higher
RMSE values due to their inability to effectively mitigate noise but achieve a sharper RMSE reduction as the SNR increases, especially in the moderate SNR
region. The AAC-OFDM waveform utilizes only the chirp rate \( \beta\) at the receiver side, while preserving all data bits for communication purposes. As a result, it exhibits a performance degradation of approximately 5 dB compared to other waveforms. However, both CM-OFDM and standard OFDM harness all symbols as pilot symbols, using the waveform solely for sensing purposes without carrying any communication data. Fig.~\ref{fig:range_rmse}(b) shows the RMSE performance when the chirp is applied across an entire slot of multiple symbols. The results show a significant improvement in RMSE compared to the single symbol case for all schemes, owing to the noise-averaging effect of multiple symbols. Both pure Chirp and CM-OFDM achieve the lowest RMSE, while AAC-OFDM also benefits from the slot-based approach, showing notable improvements in the moderate and low SNR regions. Fig.~\ref{fig:range_rmse}(c) compares the performance of CM-OFDM and OFDM using only 10\% and 25\% of pilot subcarriers for sensing instead of using all the subcarriers. Our proposed AAC-OFDM system, which performs sensing without relying on a pilot subcarrier and outperforms other waveforms, is an ideal candidate for simultaneously implementing both sensing and communication functionalities.
\begin{figure}
    \centering
    \includegraphics[scale=0.38]{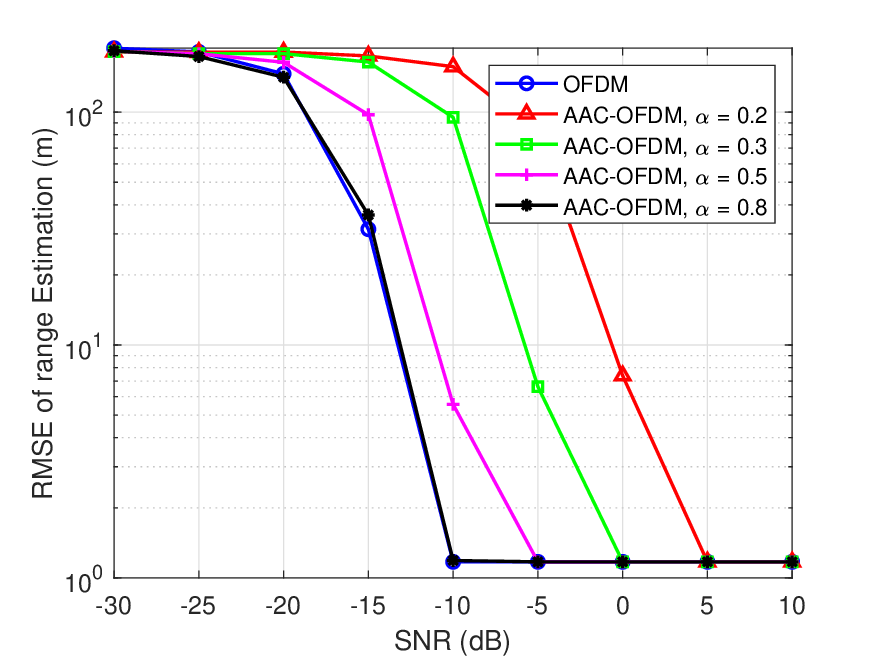} % No need to specify the full path
    \caption{Range RMSE versus SNR when chirp is introduced for one entire symbol for different $\alpha$ values.}
    \label{fig:cor_rmse_alpha_symbol}
\end{figure}
\begin{figure}
    \centering
    \includegraphics[scale=0.4]{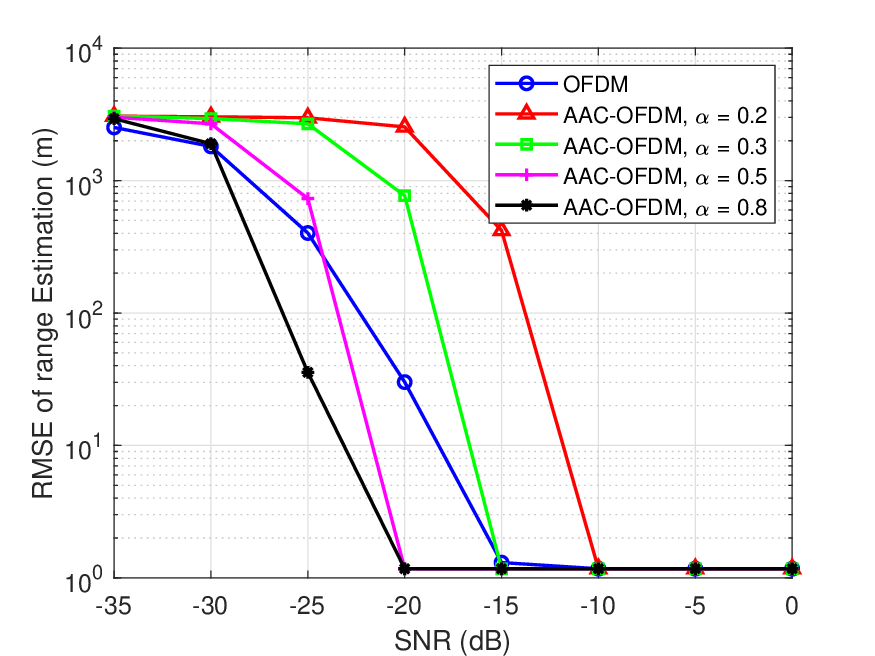} % No need to specify the full path
    \caption{Range RMSE versus SNR when chirp is introduced for one entire slot. Here is the comparison of AAC-OFDM for different $\alpha$ values. }
\label{fig:cor_rmse_alpha_slot}
\end{figure}
Fig.~\ref{fig:cor_rmse_alpha_symbol} and Fig.~\ref{fig:cor_rmse_alpha_slot} show the RMSE of the range estimation, comparing OFDM with AAC-OFDM for different weighting factors (\(\alpha = 0.2, 0.3, 0.5, 0.8\)). In Fig.~\ref{fig:cor_rmse_alpha_symbol}, the weighting factor \(\alpha\) significantly influences the RMSE performance of AAC-OFDM as compared to OFDM. For example, for \(\alpha = 0.2\) there is a notable 15 dB  SNR loss relative to OFDM, while \(\alpha = 0.5\) reduces this loss to approximately 5 dB. However, increasing \(\alpha\) degrades the BER, indicating a trade-off between sensing accuracy and communication reliability. To strike a compelling trade-off in both domains, the value of \(\alpha\) must be carefully optimized. This highlights the critical role of parameter tuning in hybrid waveforms like AAC-OFDM for ensuring effective integration of sensing and communication functionalities. 

Fig.~\ref{fig:cor_rmse_alpha_slot}, characterizes slot-wise incorporation. Observe that the performance of AAC-OFDM improves overall for all values \(\alpha\) due to the averaging effect across multiple symbols. Notably, for \(\alpha=0.8\) and \(\alpha=0.5\), AAC-OFDM outperforms standard OFDM in terms of its RMSE performance, highlighting the benefits of a stronger chirp contribution in the slot-based approach. For example, for \(\alpha=0.3\) the performance becomes comparable to that of OFDM, but for \(\alpha=0.2\) it is still unable to outperform OFDM. Despite this, the overall improvement across all configurations underscores the efficiency of slot-based symbol diversity in enhancing the sensing performance of AAC-OFDM.
\begin{figure}
    \centering
    \includegraphics[scale=0.4]{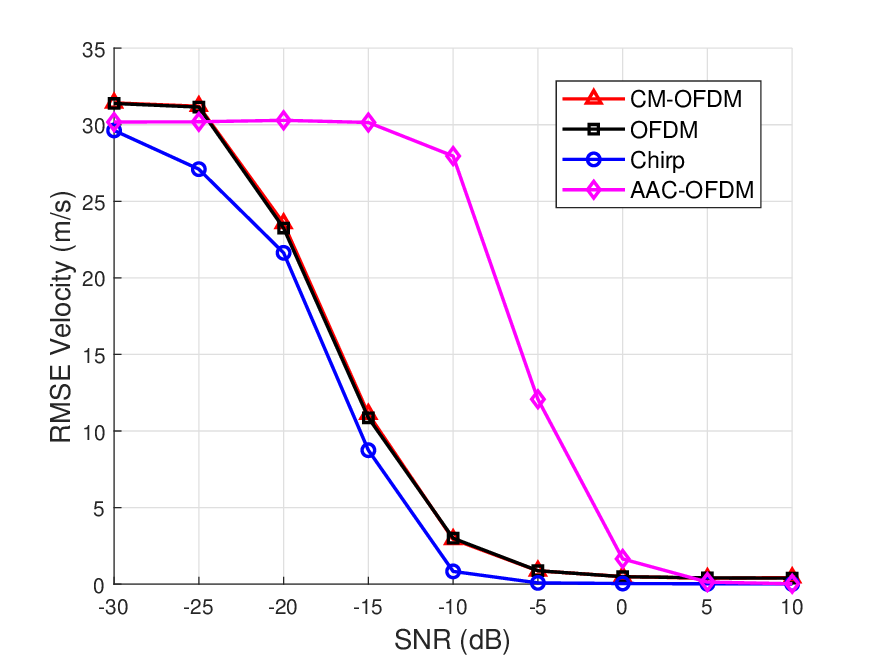} % No need to specify the full path
    \caption{Comparison of the RMSE of the velocity for the different waveforms.}
    \label{fig:velcoity_rmse}
\end{figure}
\begin{figure}
    \centering
    \includegraphics[scale=0.4]{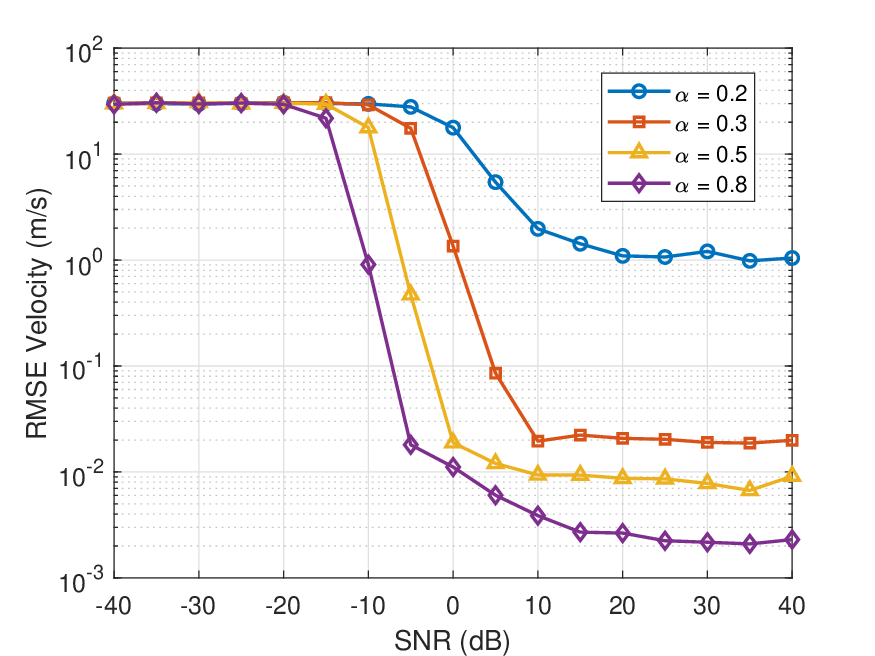} % No need to specify the full path
    \caption{ Comparison of AAC-OFDM velocity RMSE versus SNR for different $\alpha$ values.}
    \label{fig:velcoity_alpha_rms}
\end{figure}

In Fig.~\ref{fig:velcoity_rmse} we compare the RMSE of the velocity estimate of the four waveforms. The performance of the chirp waveform is superior to all the waveforms, while the velocity resolution of the AAC-OFDM and chirp was better than that of OFDM and CM-OFDM. The curves follow the estimator each receiver uses as a phase reference. At high SNR, the velocity error floors are about 0.011 m/s for chirp, 0.02 m/s for AAC-OFDM, and 0.38–0.40 m/s for OFDM and CM-OFDM. The small gap between chirp and AAC-OFDM is expected, because AAC uses a data-independent chirp template across pulses, which keeps the inter-pulse phase slope almost as clean as pure chirp, with a slight penalty from residual OFDM leakage into the chirp path that slightly perturbs sensing. Overall, AAC-OFDM reaches near-chirp Doppler accuracy around $0.02\,\mathrm{m/s}$ while still carrying data, and it clearly outperforms OFDM and CM-OFDM once the SNR is moderate to high.
 Fig. \ref{fig:velcoity_alpha_rms} shows the effect of the alpha factor on the RMSE of the velocity estimate. Here $\alpha$ is the transmit power allocated to the chirp component that the AAC correlator uses for sensing. Increasing $\alpha$ strengthens the coherent component seen by the matched filter, increases the effective sensing SNR, and reduces Doppler variance. For $\alpha=0.8$ the transition from noise-limited to estimation-limited occurs near $-12$ dB and the high-SNR floor is about $3\times 10^{-3}$ m/s. With $\alpha=0.5$ the transition is around $-5$ dB and the floor is about $1\times 10^{-2}$ m/s. With $\alpha=0.3$ the transition appears near $+5$ dB and the floor settles around $2\times 10^{-2}$ to $3\times 10^{-2}$ m/s. With $\alpha= 0.2$ the curve remains largely noise-limited even at high SNR and the velocity error stays close to one metre per second. Intuitively, the OFDM part acts as self-interference for the chirp matched filter, so allocating more power to the chirp produces a cleaner inter-pulse phase progression and a lower error floor. From a design standpoint, $\alpha$ between 0.3 and 0.5 achieves sub-0.02 m/s accuracy with balanced data performance, while $\alpha$ near 0.8 is preferable when sensing accuracy is the priority.
 
 AAC-OFDM is most suitable for opportunistic and bistatic sensing, since the $\alpha$ parameter provides a simple control to the sensing vs. communications trade-off. A high $\alpha$ can be used during initial detection to maximize sensing performance, 
which can then be reduced for steady tracking to improve data rate and BER. The alpha value can be increased again if the
SNR or sensing performance degrades. By contrast, CM-OFDM is better suited as a drop-in enhancement for existing OFDM systems, as it strengthens range resolution using known data and PRS without requiring any $\alpha$ tuning or major architectural changes.
 \vspace{-0.6em}

\section{Conclusion and Future Work}\label{sec:conclusion}
 This work introduced AAC-OFDM, a novel ISAC waveform created by affinely adding a chirp to OFDM symbols, enabling pilot-free sensing while remaining compatible with the NR resource grid. We analysed its ambiguity characteristics and sensing/communications trade-offs. Our formulation of the chirp multiplication of OFDM specifies the chirp so it fits seamlessly within the NR resource grid, i.e., it can be inserted as a PRS or standard OFDM block without affecting neighboring resource elements, providing a compatibility-first path to improved sensing. This formulation allows for the enhancement of existing PRS/OFDM procedures without affecting other resource elements. We showed that slot-level chirp placement tightens range/velocity RMSE through longer coherent integration, whereas symbol-level placement expands the unambiguous velocity window, thereby offering a practical way to balance resolution against Doppler tolerance. On the communications side, AAC-OFDM can maintain BER close to that of OFDM for modest chirp weights and can exhibit favourable PAPR trends, while preserving OFDM-like complexity.
   
    We will extend AAC-OFDM to MIMO, developing joint angle, range, and Doppler processing with transmit and receive beamforming in both single-user and multiuser MIMO, and then evaluate the resulting sensing and communications trade-offs. We will study robustness to hardware impairments such as carrier frequency offset, phase noise, and power amplifier nonlinearity, and relate their impact on BER and sensing RMSE to chirp weight tuning. Finally, we will address multi-target scenarios by improving detection and tracking so that each new measurement is matched to the correct target and its path is kept steady over time in cluttered scenes.
\vspace{-0.5em}
\bibliographystyle{IEEEtran}
\bibliography{MyRef}
\end{document}